\documentclass[fleqn,usenatbib]{mnras}
\usepackage[T1]{fontenc}
\usepackage{ae,aecompl}

\usepackage[normalem]{ulem}

\usepackage{amsmath,amssymb}
\usepackage{mathrsfs}   
\usepackage{graphicx}

\newcommand{\Msun}{M_\odot}

\newcommand{\rvir}{r_{\rm vir}}
\newcommand{\Vvir}{V_{\rm vir}}

\newcommand{\jh}{j_{\rm h}}
\newcommand{\Mh}{M_{\rm h}}
\newcommand{\Vmax}{V_{\rm max}}
\newcommand{\Vrot}{V_{\rm rot}}

\newcommand{\sigmafj}{\sigma_{f_j}}
\newcommand{\Mstar}{M_{\rm peak}}
\newcommand{\fjzero}{f_{j,0}}

\defcitealias{RF12}{RF12}

\date{December 5$^{\rm th}$ 2017}
\pubyear{2017}

\begin{document}

\title[Stellar-to-halo specific angular momentum]
{Galaxy spin as a formation probe: the stellar-to-halo specific angular momentum relation}
\author[L. Posti et al.]{Lorenzo Posti$^{1}$\thanks{E-mail: posti@astro.rug.nl},
Gabriele Pezzulli$^{2}$,
Filippo Fraternali$^{3,1}$ and
Enrico M. Di Teodoro$^{4}$
\\ \\
$^{1}$Kapteyn Astronomical Institute, University of Groningen, P.O. Box 800, 9700 AV Groningen, the Netherlands \\
$^{2}$Department of Physics, ETH Zurich, Wolfgang-Pauli-Strasse 27, 8093 Zurich, Switzerland \\
$^{3}$Dipartimento di Fisica e Astronomia, Universit\`a di Bologna, via Gobetti 93/2, I-40129 Bologna, Italy \\
$^{4}$Research School of Astronomy and Astrophysics - The Australian National University, Canberra, ACT, 2611, Australia}

\maketitle

\begin{abstract}
We derive the stellar-to-halo specific angular momentum relation (SHSAMR) of galaxies at
$z=0$ by combining i) the standard $\Lambda$CDM tidal torque theory ii) the observed relation
between stellar mass and specific angular momentum (Fall relation) and iii) various
determinations of the stellar-to-halo mass relation (SHMR).
We find that the ratio $f_j = j_\ast/\jh$ of the specific angular momentum of stars to that
of the dark matter i) varies with mass as a double power-law, ii) it always has a peak in the
mass range explored and iii) it is $3-5$ times larger for spirals than for ellipticals.
The results have some dependence on the adopted SHMR and we provide fitting formulae
in each case. For any choice of the SHMR, the peak of $f_j$ occurs at the
same mass where the stellar-to-halo mass ratio $f_\ast = M_\ast/\Mh$ has a maximum.
This is mostly driven by the straightness and tightness of the Fall relation, which requires
$f_j$ and $f_\ast$ to be correlated with each other roughly as $f_j\propto f_\ast^{2/3}$,
as expected if the outer and more angular momentum rich parts of a halo failed to accrete onto
the central galaxy and form stars (biased collapse).
We also confirm that the difference in the angular momentum of spirals and
ellipticals at a given mass is too large to be ascribed only to different spins of the parent
dark-matter haloes (spin bias).

\end{abstract}

\section{Introduction}
\label{sec:intro}

Mass and angular momentum are amongst the most important quantities  underlying a physically
motivated classification of galaxies. Inspired by this belief, \cite{Fall83} first looked at how
galaxies were distributed in the plane $j_\ast - M_\ast$ of specific angular momentum and stellar
mass. This pioneering work \citep[based on data from][] {Rubin+1980,Davies+1983} led to the discovery
of an important scaling relation, which we can now call the \emph{Fall relation(s)}: i) for a given
morphological type, stellar mass and specific angular momentum of galaxies are related by a power-law,
ii) spirals and ellipticals follow parallel sequences, with identical slopes but different
normalizations, with ellipticals having approximately five times less specific angular momentum
than spirals for a given stellar mass.

These findings were more recently confirmed by \citet[][hereafter \citetalias{RF12}]{RF12}, who
compiled measurements for a sample of $\sim 100$ nearby galaxies of all Hubble types for over three
orders of magnitude in stellar mass, by \cite{ObreschkowGlazebrook14}, who measured angular momenta
of stars, cold atomic and molecular gas in 16 nearby spirals and by \cite{Cortese+2016}, who extended
the sample to $\sim 500$ nearby galaxies and studied both the stellar and warm gas specific angular
momentum (though their measurements reached only one or two effective radii). Perhaps even more
noteworthy is that the specific angular momentum-mass relation was recently measured also for
star-forming disc galaxies at high redshift ($z\sim 1-2$) and it was found to be a power-law of
similar slope and normalization as at $z=0$ with the residuals correlating to galaxy morphology,
as observed for nearby galaxies \citep[][]{Burkert+2016,Harrison+2017}. Remarkably, the simple
nature of the Fall relation (a power-law over several decades, with little scatter) is virtually
identical to a theoretical relation which is expected to hold for the dark matter haloes from Tidal
Torque Theory \citep[TTT, see e.g.][]{Peebles69, EfstathiouJones1979}.

The simple correspondence, in the specific angular momentum-mass diagram, between
stars and dark matter could be interpreted as a major theoretical success, if one assumed
that both the mass and the angular momentum of a galaxy are directly proportional to those
of their haloes (with some dependence on morphology, but none on the mass).
Indeed, some of the earliest theoretical models of galaxy formation \citep[e.g.][]
{FallEfstathiou1980,Dalcanton+1997,MMW98} were based on these assumptions and turned out
to be relatively successful in reproducing the observed structural properties and scaling
relations of disc and spheroidal galaxies \citep[for more recent similar models see e.g.][]
{Somerville+2008,DuttonvandenBosch2012,Kravtsov13}.

We now know, however, that at least one of the two assumptions above, i.e. that the
stellar mass-to-halo mass ratio is constant with mass, is not correct.
The stellar-to-halo mass relation (SHMR) is indeed rather complex and far from being
linear as it emerged from several recent works, despite the different techniques used
to estimate dark-matter halo masses. Consistent results were found in studies that made
use of either the so-called abundance matching ansatz \citep[see e.g.][]
{ValeOstriker2004,Moster+2013,Behroozi+2013}, sometimes combined with direct estimates
via weak lensing and stellar or satellite kinematics \citep[see e.g.][]{Leauthaud+2012,
vanUitert+16}, or the spatial clustering of galaxies \citep[see e.g.][]
{Yang+2003,Tinker+2017} or the segregation by colour of galaxy populations
\citep[see e.g.][]{Dutton+2010,RodriguezPuebla+2015}.
In the light of these findings, the simplicity of the observed Fall relation ceases to
be a confirmation of expectations and rather starts to be a \emph{theoretical challenge}.
In particular, the following question naturally arises:
what fraction of the specific angular momentum of haloes their galaxies need
to have to be consistent simultaneously with the Fall relation and the SHMR?
Addressing this question is the main aim of this paper.

From the observational point of view, only very few studies have tried to investigate
somewhat similar questions: for instance, \cite{Kauffmann+2015} have measured the specific
angular momenta of $\sim 200$ massive gas rich galaxies from rotation curves,
they have computed the fraction of stellar-to-halo specific angular momentum and found it
to strongly correlate with galaxy mass. From the theoretical point of view, some
models were already built upon the realisation that if the SHMR is not constant with
mass, then galaxies should have a non-constant fraction of the specific
angular momentum of their haloes \citep[e.g.][]{NavarroSteinmetz2000}.
More recently, also other authors have investigated similar questions to ours by means
of semi-analytical models \citep[e.g.][]{DuttonvandenBosch2012,Stevens+2016}, as well as
of cosmological simulations
\citep[][]{Genel+2015,PedrosaTissera2015,Teklu+2015,Zavala+2016,Sokolowska+2017}.
Despite going deep in physically-motivated prescriptions, all these studies either
did not really try or fell short of reproducing the straightness and the small
scatter of the Fall relation across its entire mass range and many did not really
compare the predictions of their models on the specific angular momentum-mass
diagram and on the SHMR at the same time, hence leaving unanswered questions on the
galaxy-halo connection.

In the present contribution, we adopt a complementary approach. Without assuming
a specific scenario, we try to reconstruct the stellar-to-halo specific angular
momentum relation (SHSAMR), as a function of mass and morphological type, by requiring
the empirical Fall relation to be reproduced. We stress the inherent uncertainties of
the process by systematically investigating the dependence of the results on the adopted
SHMR. Then, we try to address some interesting related questions, such as i) whether
the origin of the specific angular momentum-mass relation can be
interpreted in terms of galaxies forming stars from the innermost and more angular
momentum-poor part of the halo and ii) whether the different angular momenta between
spirals and ellipticals can be ascribed to differences in the spin parameters of the
host dark matter haloes.

The paper is organised as follows. In Section \ref{sec:model}, we briefly review some
observational and theoretical background concerning the angular momentum of the stellar
and dark components of galaxies. In Section \ref{sec:ret_frac}, we derive our SHSAMR
for spirals and ellipticals, adopting different existing SHMR from the literature.
In Section \ref{sec:forward_models}, we investigate the origin of the  Fall relation
with two physical models (biased collapse and spin bias).
We summarize and conclude in Section \ref{sec:concl}.

Throughout the paper we use the standard $\Lambda$ Cold Dark Matter ($\Lambda$CDM)
model with the cosmological parameters estimated by the \cite{Planck14}:
$(\Omega_{\rm m}, \Omega_{\Lambda}) = (0.31, 0.69)$ are the matter and dark energy
density and $H_0=67.7 \, {\rm km\,s^{-1}\,Mpc^{-1}}$ is the Hubble constant.

\section{The angular momentum of galaxies and their haloes}
\label{sec:model}

Here we introduce the framework with the equations relevant to our
work. We begin by re-deriving the specific angular momentum of dark matter haloes in
a standard $\Lambda$CDM Universe and then we relate this to galaxies by introducing
the key parameters of our models.

\subsection{The specific angular momentum of dark matter haloes}
\label{sec:lcdm}

The mass distribution of a spherically symmetric dark matter halo whose density distribution
is described by a \citet[][hereafter NFW]{NFW96} profile, is characterized by two
parameters: the mass within the virial radius $\Mh$ and the concentration $c$.

At $z=0$, the virial radius of a halo of mass $\Mh$ is
\begin{equation}\label{def:rvir}
\rvir = \left( \frac{2G\Mh}{\Delta_{\rm c} H_0^2} \right)^{1/3}
\end{equation}
where $\Delta_{\rm c}=102.5$ is the virial overdensity w.r.t. to the critical
density of the Universe \citep[][]{BryanNorman1998}.

The mass-concentration relation has been widely studied in the literature, especially by
means of cosmological simulations \citep[e.g.][]{NFW97} which have shown that the
concentration has a weak dependence on halo mass. In the following we adopt the relation
found by \cite{DM14}
\begin{equation}
\label{eq:cm}
\log\, c = 0.537 - 0.097 \log[\Mh/(10^{12}\Msun/h)]
\end{equation}
with an intrinsic scatter of about $0.11$ dex.

Dark matter haloes acquire rotation from tidal torques exerted by the surrounding matter
\citep[see e.g.][]{Hoyle1949} and this is usually quantified with the so-called spin
parameter
\begin{equation}
\label{def:lambda}
\lambda \equiv \frac{J\,|E|^{1/2}}{G\,M^{5/2}},
\end{equation}
where $G$ is the gravitational constant and $J$, $M$ and $E$ are respectively the
angular momentum, mass and energy of the system \citep[see][]{Peebles69}.
Cosmological simulations have shown that $\lambda$ is log-normally distributed and
independent on halo mass \citep[e.g.][]{EfstathiouJones1979,BarnesEfstathiou1987,
Porciani+2002}. Here we adopt the results of \cite{Maccio+2007}, who found log-normal
distributions with expectation value $\lambda \simeq 0.035$ and scatter of about $0.25$ dex.

By inverting the definition of $\lambda$ (equation \ref{def:lambda}), the specific
angular momentum $\jh$ of a dark matter halo can be conveniently expressed as
\begin{equation}
\label{eq:j_halo}
\jh \equiv \frac{J_{\rm h}}{\Mh} = \sqrt{\frac{2}{F_E(c)}} \lambda \Vvir \rvir,
\end{equation}
where
\begin{equation}\label{def:Vvir}
\Vvir\equiv\sqrt{\frac{G\Mh}{\rvir}}
\end{equation}
is the halo circular velocity at the virial radius and
\begin{equation}\label{def:F_E}
F_E \equiv -\frac{2E}{\Mh \Vvir^2} \;.
\end{equation}
is a dimensionless factor that depends on the structure of the halo. Following
\citet[][their equation 23] {MMW98}, we write $F_E$ as a function of the concentration
$c$:
\begin{equation}
\label{eq:F_E}
F_E(c) = \frac{c}{2}\frac{\left[1-1/(1+c)^2-2\ln(1+c)/(1+c) \right]}
{\left[c/(1+c)-\ln(1+c)\right]^2}.
\end{equation}

Combining equations \eqref{def:rvir}, \eqref{eq:j_halo} and \eqref{def:Vvir} we find
that at $z=0$ the specific angular momentum $\jh$ of a halo of mass $\Mh$ is
\begin{equation}
\label{eq:jh-Mh}
\left. \begin{array}{lll}
\jh & = &(\Delta_{\rm c}\,H_0^2)^{-1/6} \dfrac{\lambda}{\sqrt{F_E(c)}}
	\left( 2 G \Mh  \right)^{2/3} \\
&& \\
& = & \dfrac{1.67 \times 10^3}{\sqrt{F_E(c)}} \left( \dfrac{\lambda}{0.035} \right)
	\left( \dfrac{\Mh}{10^{12} \Msun} \right)^{2/3} \; {\rm kpc\; km\; s^{-1}},
\end{array}\right.
\end{equation}
which is not very different from $\jh\propto\Mh^{2/3}$, since the dependence of
$\sqrt{F_E(c)}$ on halo mass is very weak\footnote{
$F_E(c)$ varies about $20\%$ in the mass range $10\leq\log\Mh/\Msun\leq 14$.
} compared to the power-law $\Mh^{2/3}$.

This is similar to the relation adopted by \citetalias{RF12}, except for the fact that
we use updated cosmological parameters and we take the dependence on the concentration
into account.

\begin{figure*}
\includegraphics[width=\textwidth]{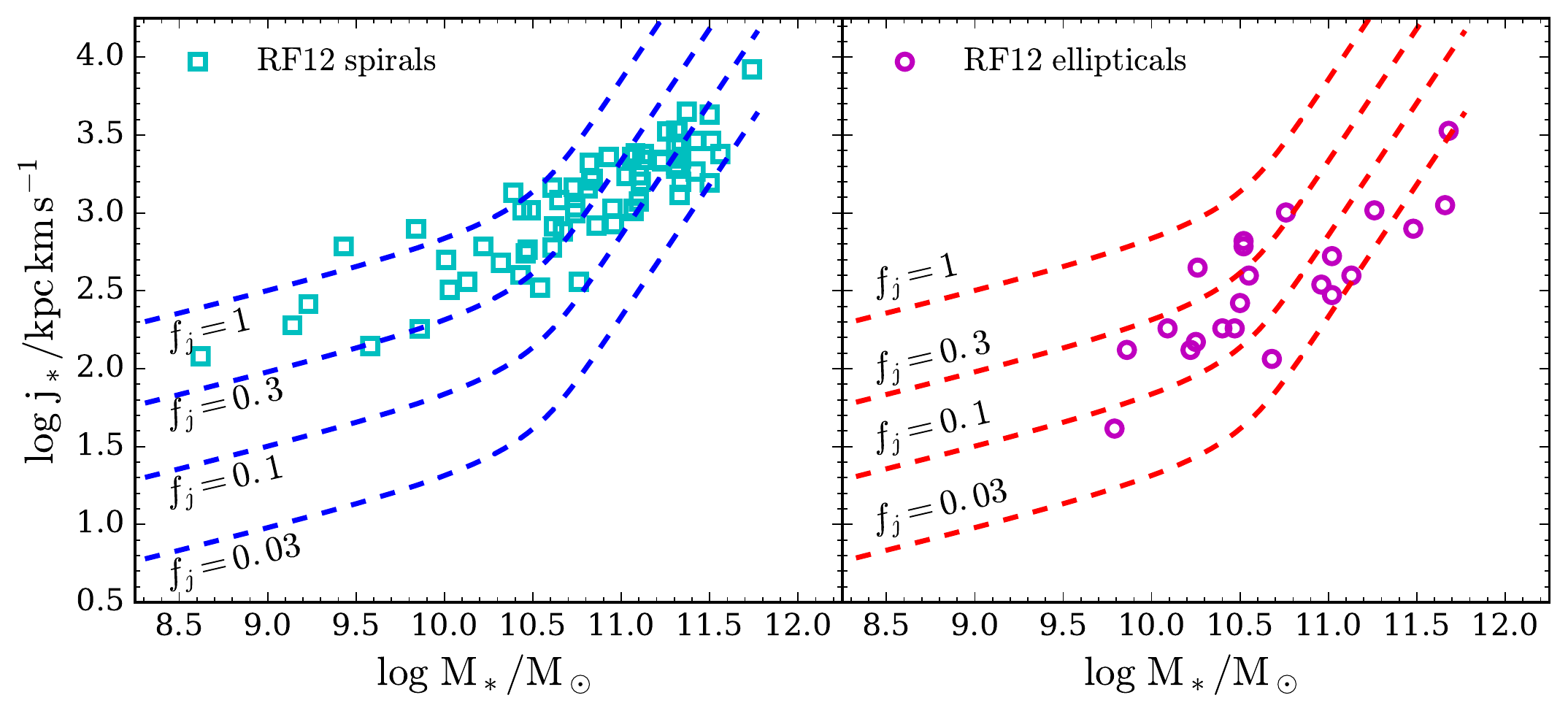}
\caption{Specific angular momentum-mass diagram for a model with the SHMR of
		 \citet[][]{Moster+2013} and a constant retained fraction $f_j$ for late and early
         types in the left- and right-hand panel respectively (assuming $\lambda=0.035$
         for the dark matter haloes).
         Different curves are for different values of the constant $f_j=1,0.3,0.1,0.03$.
         The cyan squares and the magenta circles are data points for late and early types
         respectively from \citetalias{RF12}.
}
\label{fig:shmr_rf12}
\end{figure*}

\subsection{The specific angular momentum of the stars}
\label{sec:jstar}

From equation \eqref{eq:jh-Mh}, the average specific angular momentum of the stars in a
galaxy can be written as
\begin{equation}\label{jstarMstar_theory}
j_\star = \frac{ 77.4 }{\sqrt{F_E(c)}} \left( \frac{\lambda}{0.035} \right)
	f_j f_\star^{-2/3} \left( \frac{M_\star}{10^{10} M_\star} \right)^{2/3} \;
    {\rm kpc\; km\; s^{-1}},
\end{equation}
where
\begin{equation}
f_\star \equiv \frac{M_\star}{\Mh}
\end{equation}
is the ratio of stellar mass to halo mass, while
\begin{equation}
\label{eq:fj}
f_j \equiv \frac{j_\star}{\jh}
\end{equation}
is the ratio of the average specific angular momentum of the stars in the galaxy to
that of the dark matter halo\footnote{Equation \eqref{jstarMstar_theory} is also similar
to the relation adopted by \citetalias{RF12}, but for the adopted cosmology, the
dependence on concentration and the fact that they normalize the stellar fraction
by the cosmic baryon fraction.}.

The ratio of stellar mass to halo mass $f_\star$ is sometimes loosely referred to
as \emph{star formation efficiency}, although it encapsulates a variety of very different
processes, including e.g. the actual efficiency of the conversion of gas into stars,
the fraction of baryons which ever collapses to the centre of the halo
\citep[e.g.][]{Kassin+2012}, the cumulative effect of feedback from star formation
and accretion on the central black hole \citep[e.g.][]{Veilleux+2005, Harrison2017}
and, finally, the effect of galactic fountains \citep[which can engage a significant
portion of gas in circulating in the circumgalactic medium at various distances from
the star forming body of the galaxy, e.g.][]{FraternaliBinney2008}. Notice also that,
by construction, $f_\star$ is bound to be equal or smaller than the \emph{cosmic baryon fraction}
$f_{\rm b} \equiv \Omega_\textrm{b}/\Omega_\textrm{m} = 0.157$ \citep[][]{Planck14}.

Similarly, we will somewhat loosely refer to $f_j$  in equation \eqref{eq:fj} as the
\emph{retained fraction of angular momentum}. This parameter depends on many phenomena,
such as
the angular momentum losses due to, e.g., dynamical friction of baryonic massive clumps
or induced by galaxy mergers \citep[see e.g.][]{AarsethFall1980,Barnes92,HernquistMihos1995},
but also on the possibility that galaxies form from the inner and less angular momentum rich
portion of haloes \citep[\emph{biased collapse}, e.g.][\citetalias{RF12}]{vandenBosch1998},
partially counteracted by the opposite effect, namely the selective removal of low angular
momentum material as a consequence of galactic winds \citep[e.g.][]
{Governato+2007,Brook+2011},
and finally, to some extent, angular momentum redistributions due to galactic fountains
\citep[e.g.][]{Brook+2012,DeFelippis+2017}. Both estimates from recent observations
\citep[e.g.][]{Cortese+2016} and expectations from numerical simulations \citep[e.g.][]
{Teklu+2015} and semi-analytic models \citep[e.g.][]{DuttonvandenBosch2012} agree that this
fraction should be of the order of $f_j \sim 0.1-0.5$ for all galaxies, even if in principle
such fraction can be also larger than one.

As discussed in Section \ref{sec:intro}, observational determinations of the
$j_\star-M_\star$ relation find a power-law dependence $j_\star \propto M_\star^\alpha$
with index $\alpha$ close to $2/3$ for each morphological type (the \emph{Fall relation}).
A comparison with equation \eqref{jstarMstar_theory} therefore suggests that the quantity
\begin{equation}
\label{eq:const_fjfstar}
Q = \frac{\lambda}{F_E(c)}f_j f_\star^{-2/3}
\end{equation}
does not vary much with stellar or halo mass, at least for a fixed morphological type.
Amongst the ways to have $Q$ constant there is the possibility that both $f_\ast$ and $f_j$
are constant \citep[e.g.][]{MMW98} or that $f_\ast$ and $f_j$ are correlated \citep[e.g.,][]
{NavarroSteinmetz2000}.

In contrast, any model with varying $f_\ast$ and constant $f_j$ as a function of galaxy mass
\emph{are not expected to be in agreement with the data}.
This is illustrated in Figure \ref{fig:shmr_rf12}, where we plot for spirals (left panel)
and ellipticals (right panel) the predictions on the specific angular momentum-mass
diagram of a model, given by equation \eqref{jstarMstar_theory}, with variable $f_\ast$ and
constant $f_j$. For this example, we fixed the spin and concentration of the haloes to a
typical value ($\lambda=0.035$ and $c=10$) and we use the widely adopted SHMR of
\citet{Moster+2013}. Clearly, none of the curves shown provides a satisfactory description
of the data. This indicates that $f_j$ is not constant, but rather varies with both mass and
morphological type, being close to $0.5$ for low-mass spirals and about $0.03$ for the most
massive galaxies of any type. This is true, at least, if both the Fall relation and the SHMR
of \citet{Moster+2013} are correct.

In the next Section we investigate the dependence of $f_j$ on galaxy mass and morphological
type in detail and for different choices of the SHMR.

\section{The retained fraction of angular momentum as a function of galaxy mass and type}
\label{sec:ret_frac}

\begin{figure}
\includegraphics[width=.49\textwidth]{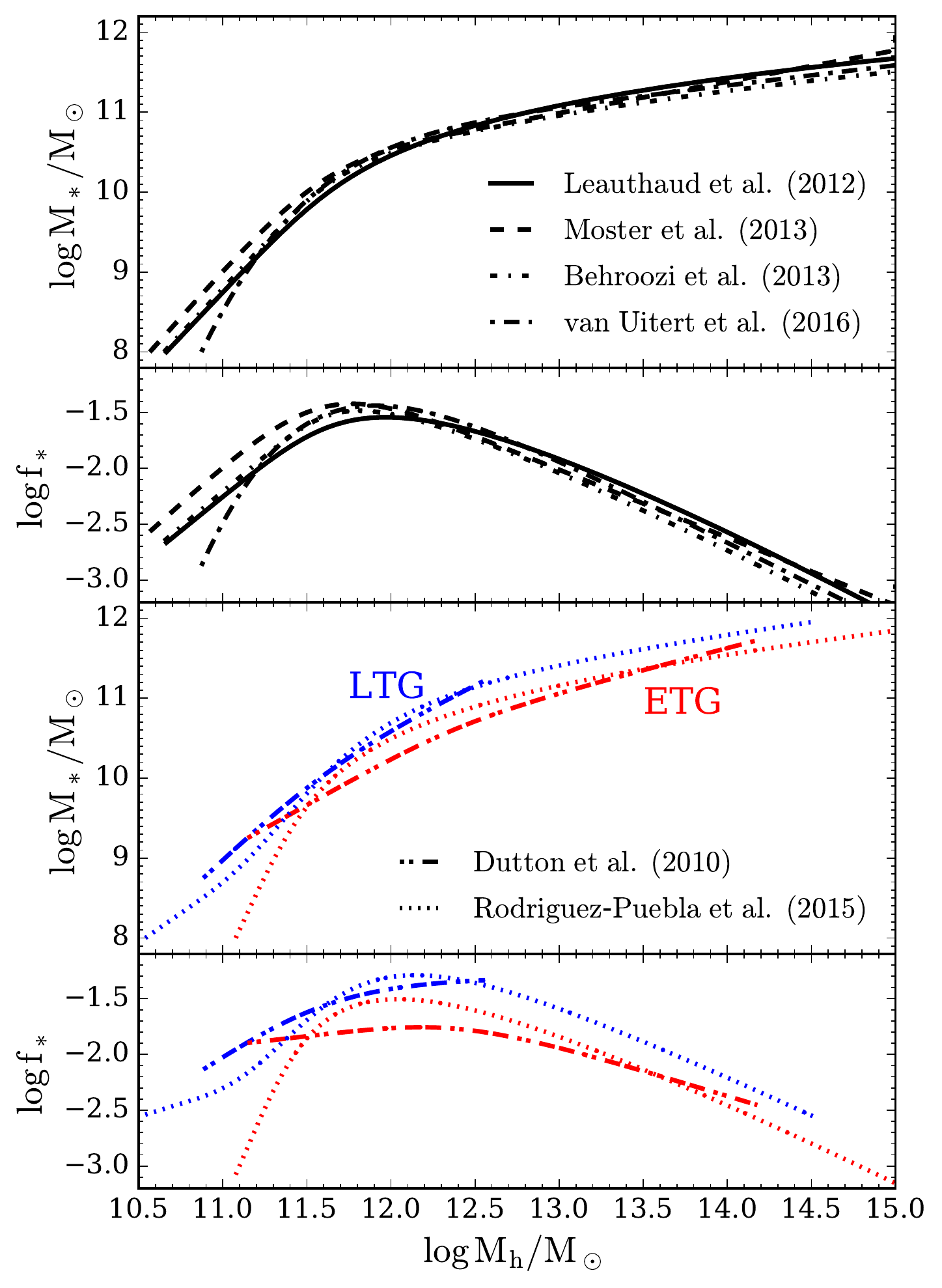}
\caption{The mean SHMR and the corresponding $f_\ast=f_\ast(\Mh)$ for the two
		 parametrisations derived for different galaxy types (bottom panels, blue for late
         types and red for early types) and the other four adopted in this work
         (top panels).}
\label{fig:shmr}
\end{figure}

\subsection{Summary of the model}
\label{sec:mod_summary}

We now derive the SHSAMR as a function of galaxy mass and morphological type by imposing that
both the Fall relation and a given SHMR are satisfied.
We proceed as follows:

\begin{itemize}
\setlength\itemsep{0.5em}
\item[(i)] we draw a sample of $10^5$ galaxies from a uniform distribution\footnote{
		   Since we are interested in the behaviour of some scaling relations and not in
           galaxy/haloes number densities, we do not need to adopt a more realistic galaxy
           mass function. We work with a uniform mass function for clarity of the
           presentation of the results.} in $\log M_\ast$
		   in the range $8.5 \leq\log (M_\ast/\Msun)\leq 12$;
\item[(ii)] to each galaxy we assign a dark matter halo whose mass is drawn from a
			normal distribution in $\log\Mh$ with mean and standard deviation given by the
            adopted SHMR (see Section \ref{sec:shmr});
\item[(iii)] we draw a spin parameter from a normal distribution in $\log\lambda$
			 with mean $\lambda=0.035$ and standard deviation of $0.25$ (see Section
             \ref{sec:lcdm});
\item[(iv)] we draw a concentration $c$ from a normal distribution in $\log\,c$ whose mean
			and standard deviation are given by the $c(\Mh)$ as in equation \eqref{eq:cm};
\item[(v)] we compute $\jh$ using equation \eqref{eq:jh-Mh};
\item[(vi)] we draw the stellar specific angular momentum $j_\ast$ from a normal distribution
			in $\log\,j_\ast$ with mean and standard deviation as estimated by
            \citetalias{RF12}. In particular we use the two following estimates of the Fall
            relation:
            \begin{equation}
            \log j_\ast = 3.18 + 0.52 \left[ \log (M_\ast/\Msun) - 11 \right],
                \; \sigma_{\log j_\ast}=0.19,
            \label{eq:jstar-Mstar_rf12_ltg}
            \end{equation}
            for late-type galaxies in the stellar mass range $8.62 \leq \log(M_\ast/\Msun)
            \leq 11.74$ and
            \begin{equation}
            \log j_\ast = 2.73 + 0.60 \left[ \log (M_\ast/\Msun) - 11 \right],
                \; \sigma_{\log j_\ast}=0.24,
            \label{eq:jstar-Mstar_rf12_etg}
            \end{equation}
            for early-type galaxies in $9.79 \leq \log(M_\ast/\Msun) \leq 11.94$, where
            $\sigma_{\log j_\ast}$ is the root mean square
            scatter. ;
\item[(vii)] we compute $f_j=j_\ast/\jh$.
\end{itemize}

\begin{figure}
\includegraphics[width=.49\textwidth]{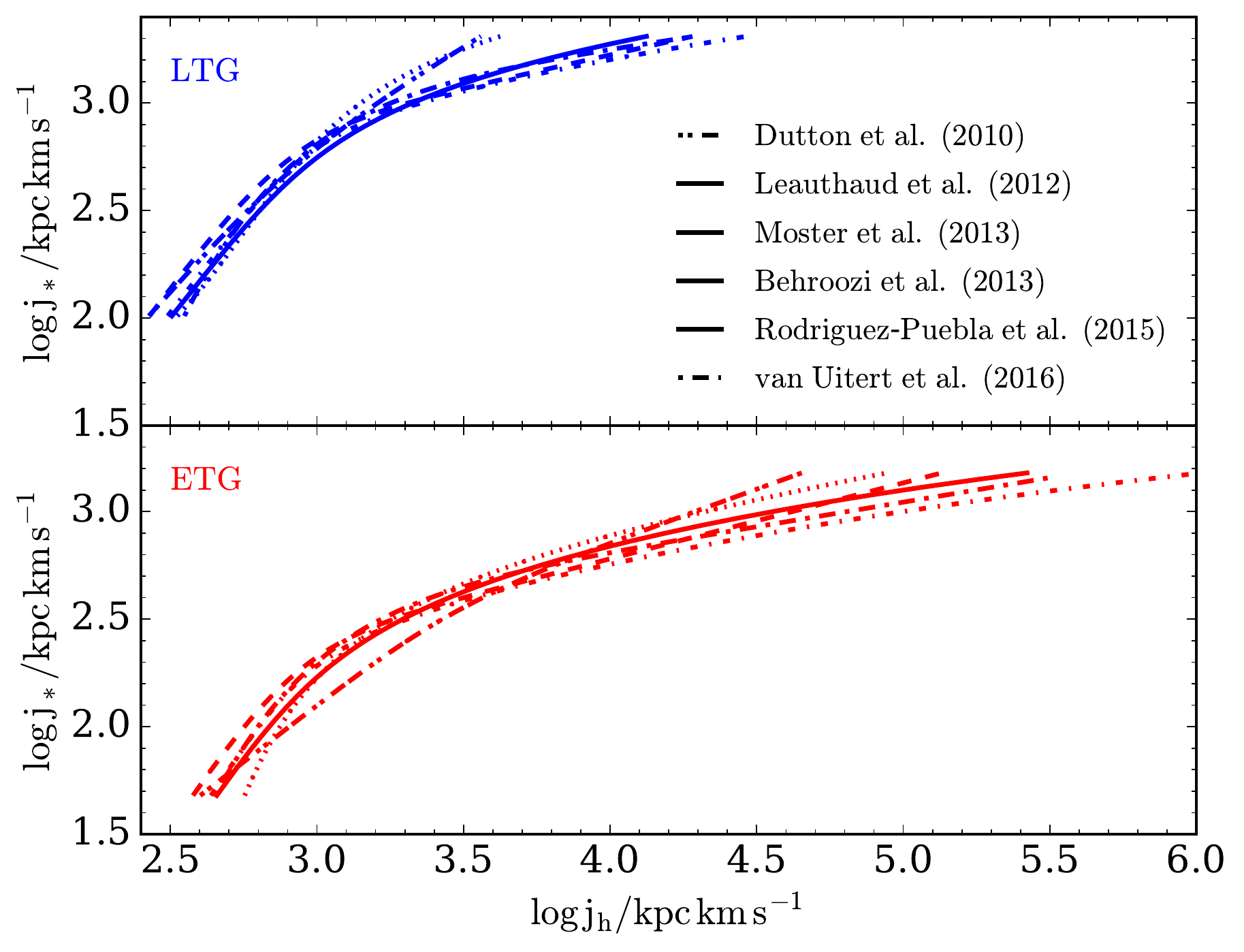}
\caption{The mean SHSAMR for late- (top panel) and early-type galaxies
		 (bottom panel) for six different SHMR. This relation is computed in the stellar
         mass ranges probed by the measurements of spirals and elliptical galaxies from
         \citetalias{RF12}.}
\label{fig:shsamr}
\end{figure}

Note that this procedure implicitly assumes that the scatters of all the adopted relations
are uncorrelated. The implications of this assumption are discussed in Section
\ref{sec:scatter}.

For simplicity, in what follows we will use the terms late-type and early-type galaxy
as synonyms for spiral and elliptical galaxy. This implies, for instance, that we are
assuming that the population of spirals and ellipticals of \citetalias{RF12} traces the
population of blue and red galaxies when we adopt colour-segregated SHMR \citep[][]
{Dutton+2010,RodriguezPuebla+2015}. Also, we do not explicitly treat intermediate galaxy
types, e.g. S0s, for which the $j_\ast-M_\ast$ relation would be intermediate between
equations \eqref{eq:jstar-Mstar_rf12_ltg} and \eqref{eq:jstar-Mstar_rf12_etg}, see
\citetalias{RF12}. However, in this work we are interested in the \emph{main trends} of
the galaxy populations as a function of morphological type and mass and we expect this
rather crude distinction to capture them.

\subsection{The SHMR}
\label{sec:shmr}

We employ six different parametrisations of the SHMR amongst those available in the
literature. We consider three types of SHMR and two examples per type.

The first two are derived by imposing that the cumulative mass function of haloes matches the
observed cumulative galaxy stellar mass function, i.e. the so-called \emph{abundance matching}
technique \citep[see e.g.,][]{ValeOstriker2004}. We adopt in particular the two very popular models
by \cite{Behroozi+2013} and \cite{Moster+2013}, which are constrained using the largest compilation
of galaxy stellar masses available (including galaxies of all types) at different redshifts.
We show the (mean) SHMR and the corresponding $f_\ast$ in Figure \ref{fig:shmr} (top panels).
The scatter of these two relations is about $\sim 0.15$ dex

The second set of SHMR we consider is derived combining constraints from the abundance matching
hypothesis together with empirical estimates of $M_\ast$ and $\Mh$ for individual objects, e.g.
from stellar, gas or satellite kinematics or from weak galaxy-galaxy lensing.
This should, in principle, lead to a better estimation of the global SHMR. Note, however, that
including individual objects in the fit may bias the results towards a given population of
galaxies (e.g. galaxies with well-behaved gas discs if one uses gas kinematics or galaxies in
crowded environments if one uses weak galaxy-galaxy lensing). In this work we consider the two
parametrizations of \cite{Leauthaud+2012} and \cite{vanUitert+16}, which we plot in Figure
\ref{fig:shmr} (top panels). The scatter of these two relations is about $0.19$ dex and $0.15$ dex
respectively.

Finally, we use two SHMR that have been estimated separately for galaxies with red and blue colours
(either with pure abundance matching or by combining different constraints) and in particular those
of \cite{Dutton+2010} and \cite{RodriguezPuebla+2015}. The first attempt in this direction was that
of \cite{Dutton+2010}. However, this study has a few limitations: i) the stellar mass range probed
is small ($9.7 \lesssim \log(M_\ast/\Msun) \lesssim 11.2$ for spirals), ii) measurements of both
abundance matching models and stellar/halo masses for individual galaxies have improved since $2010$
and iii) the best-fitting function itself is found by eye. All these reasons probably lead to a
rather flat estimate of $f_\ast$ for spirals galaxies if compared to those of many other works (see
e.g. Figure \ref{fig:shmr}). The more recent work by \cite{RodriguezPuebla+2015} improved over these
issues as  i) they probe a much larger mass range $9 \lesssim \log(M_\ast/\Msun) \lesssim 12$, ii)
they use updated data and methods to match the stellar/halo mass functions and iii) they use an
automatic routine to find the best-fitting parameters of their model. We will, however, still
consider the SHMR by \cite{Dutton+2010} in order to be able to compare our results with those of
\cite{DuttonvandenBosch2012}, who first inferred the $f_j-\Mh$ for spiral galaxies. We show these
two SHMR segregated by galaxy type in Figure \ref{fig:shmr} (bottom panels).

\begin{figure*}
\includegraphics[width=0.329\textwidth]{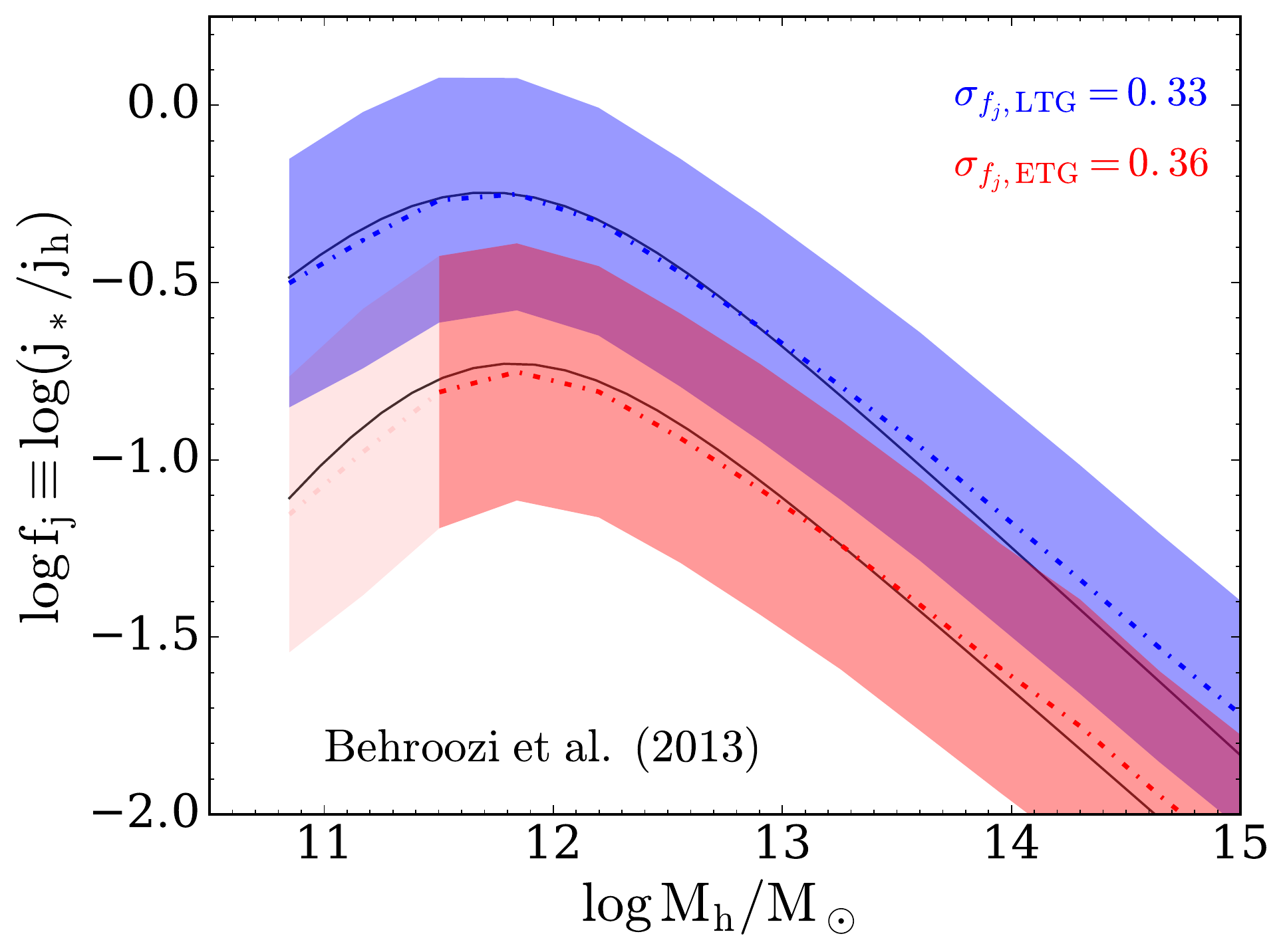}
\includegraphics[width=0.329\textwidth]{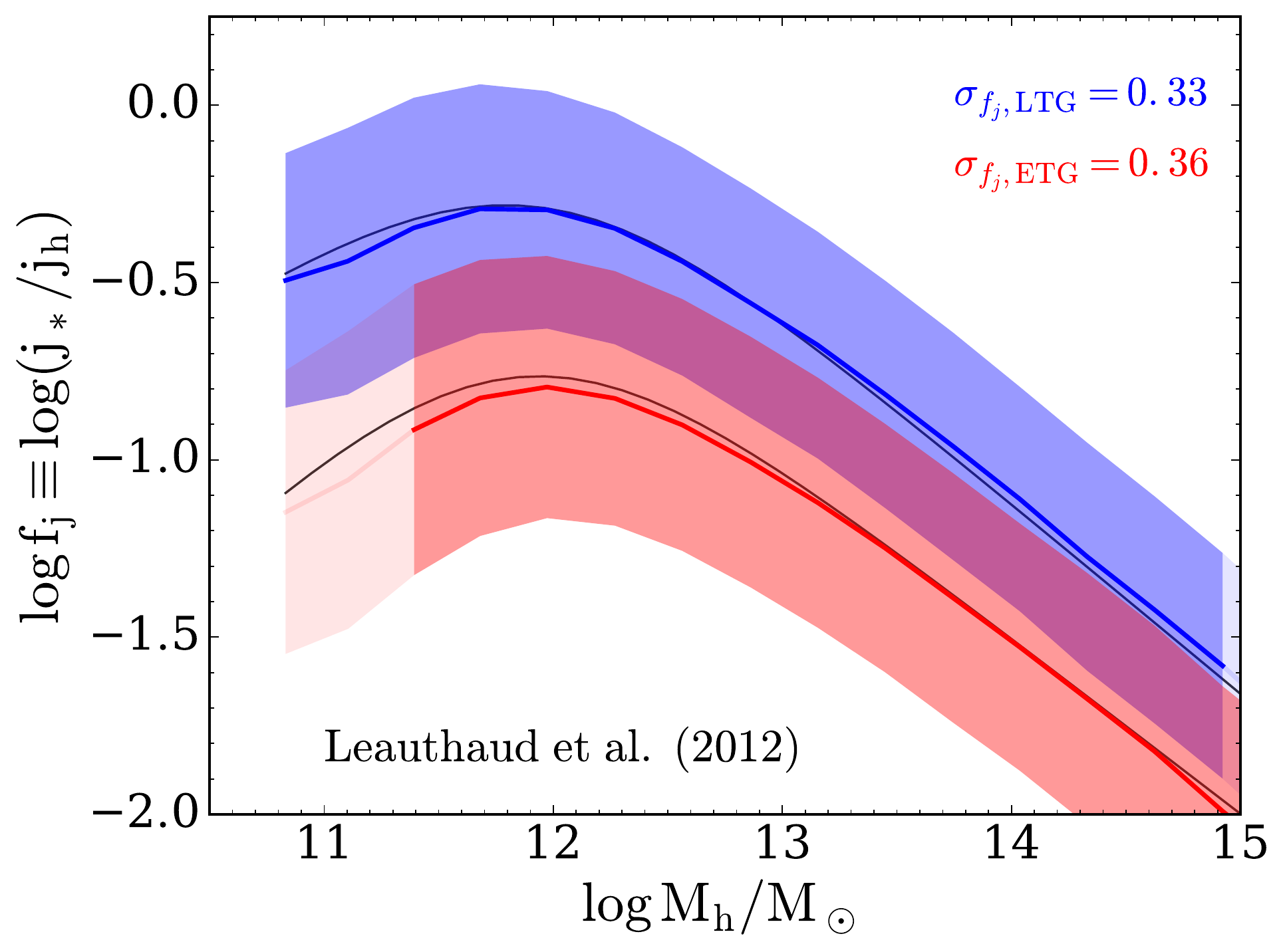}
\includegraphics[width=0.329\textwidth]{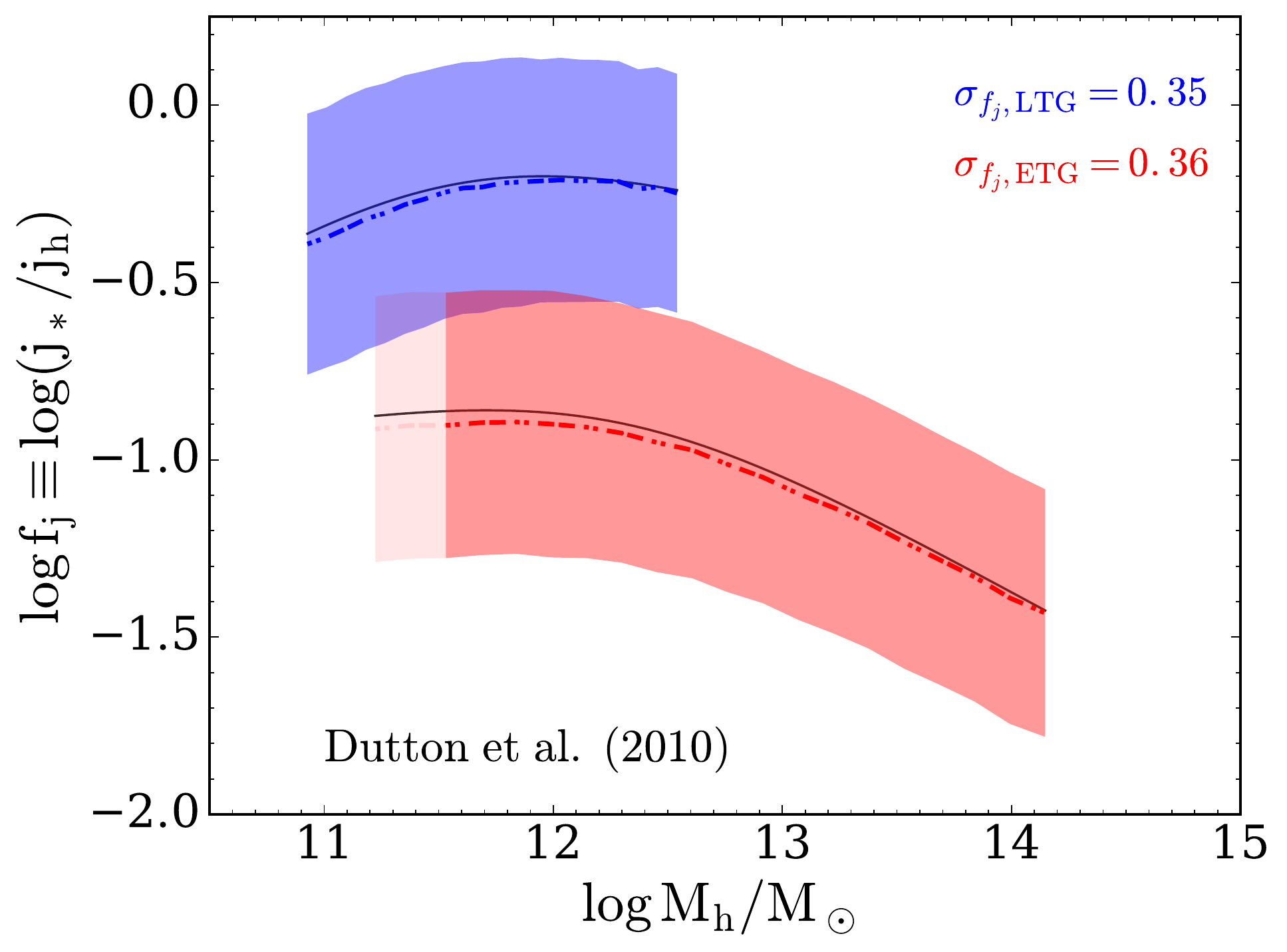} \\
\includegraphics[width=0.329\textwidth]{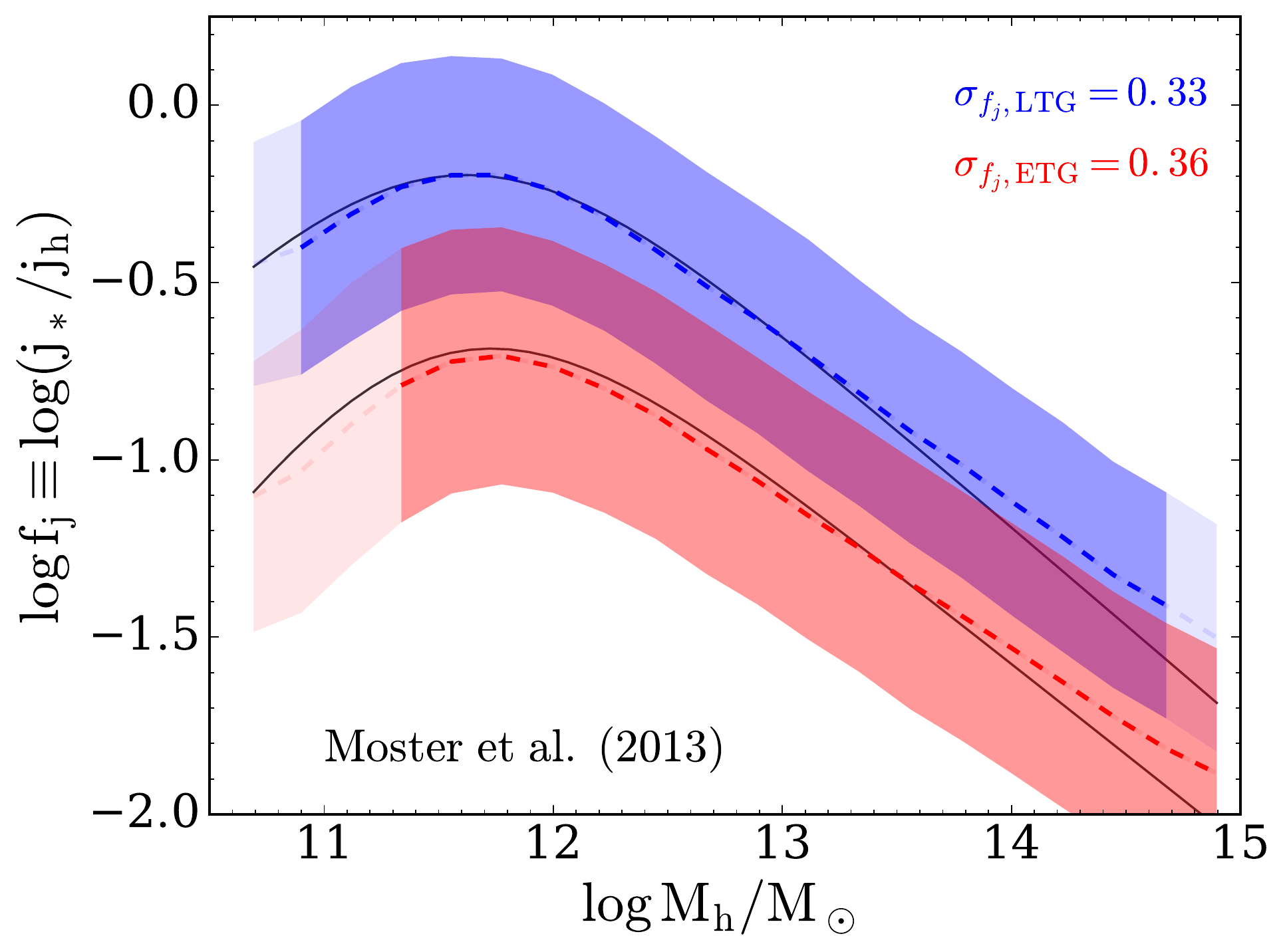}
\includegraphics[width=0.329\textwidth]{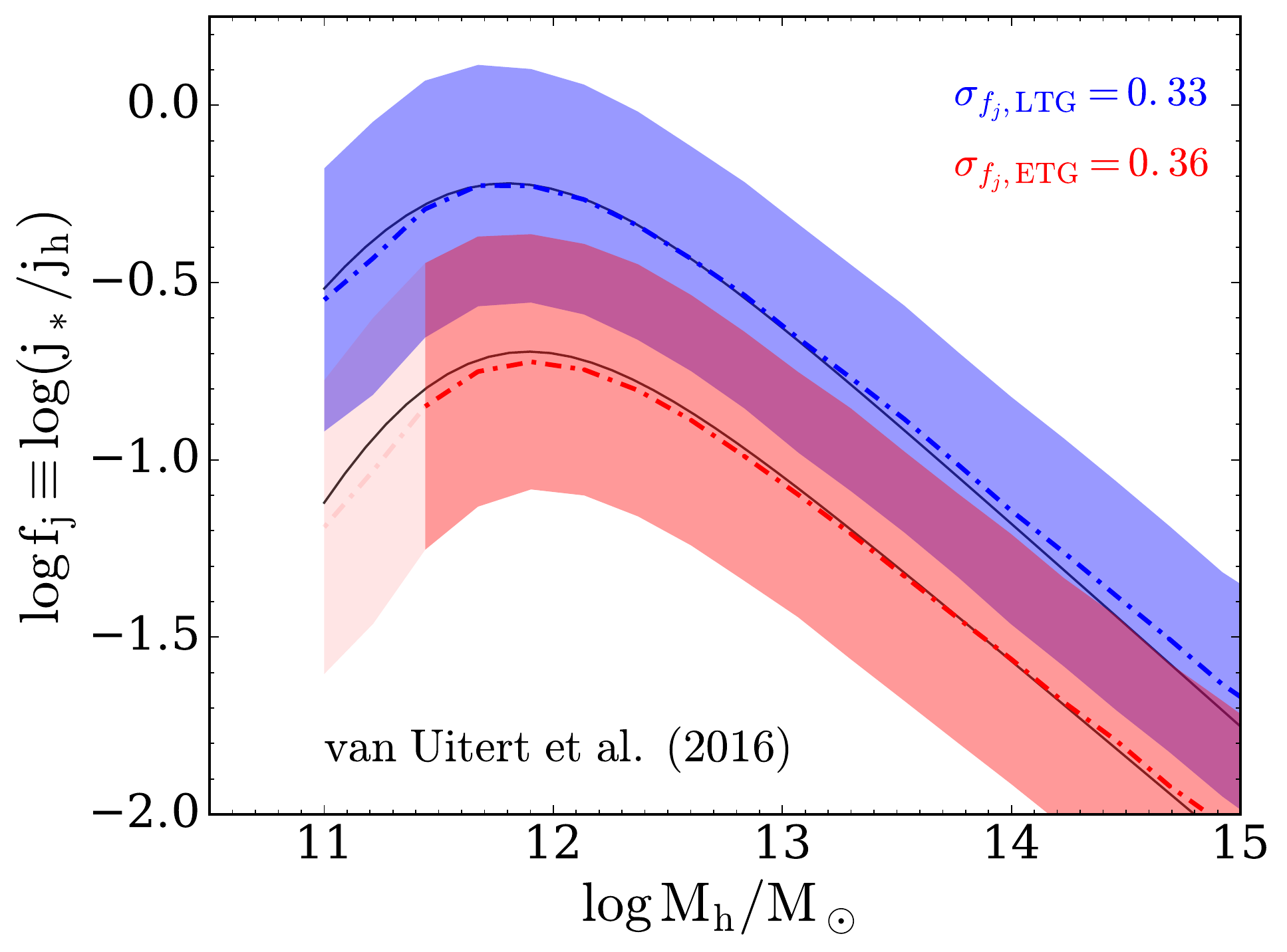}
\includegraphics[width=0.329\textwidth]{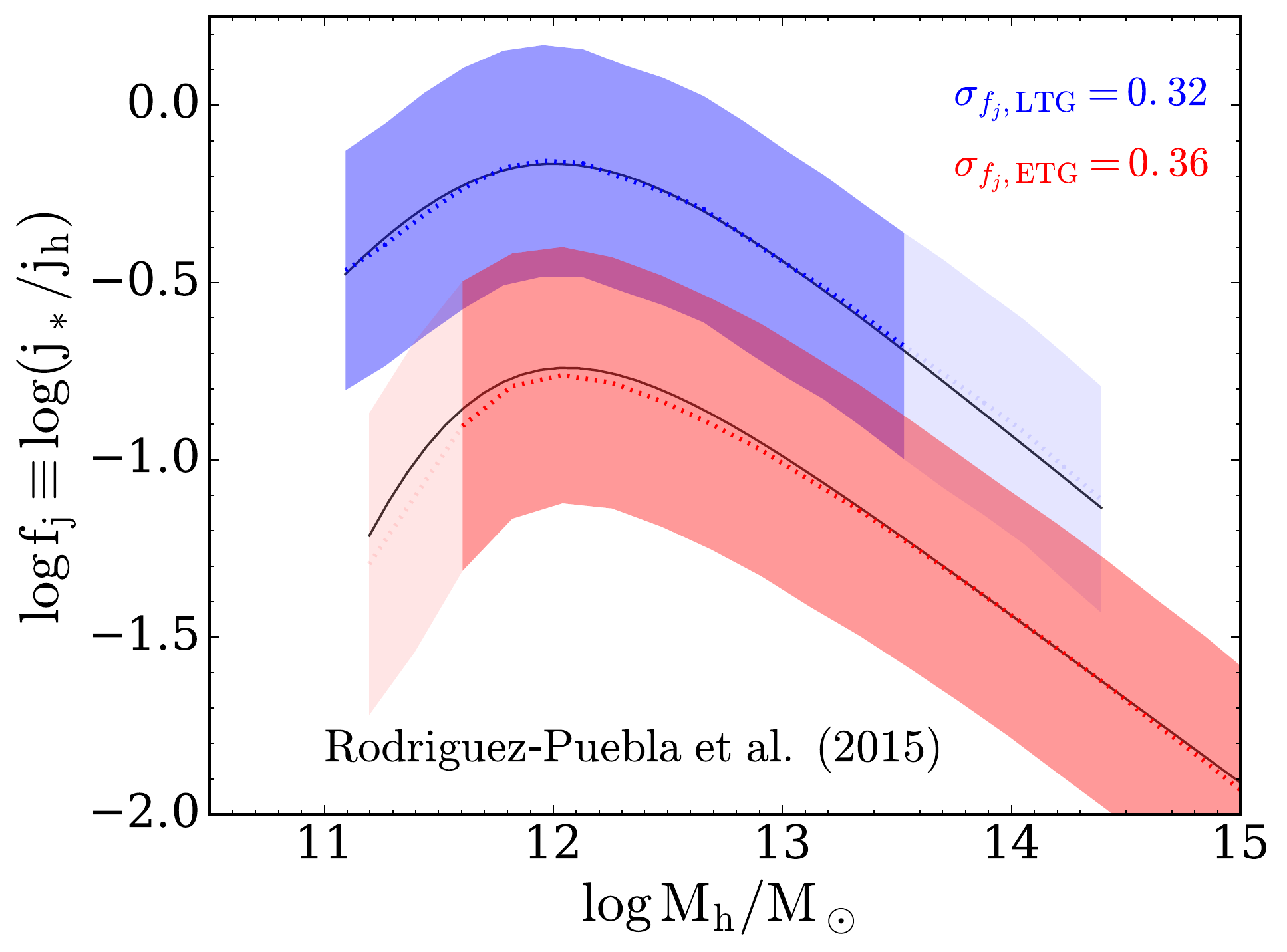}
\caption{Ratio of stellar to halo angular momentum as a function of halo mass for six models with
		 different SHMR. In all the six panels the blue and red curves (shaded regions) are the
         mean (1$\sigma$ interval) for the late-type and early-type galaxies respectively.
         The models are plotted in the fiducial galaxy mass range for the given SHMR.
         Darker shaded areas denote the mass range probed by \citetalias{RF12}. Lighter areas
         assume an extrapolation of the best fit power-law of \citetalias{RF12} beyond that range.
         The right panels are for models with different SHMR for the two galaxy types, while
         the left and central panels are for those which have the same SHMR for both types.
         In each panel we also plot, with thin black solid curves, the best-fit double power-law
         $f_j(\Mh)$ parametrized as in equation (\ref{eq:fits}). The values of the best-fit parameters
         are given in Table \ref{tab:fits}. In the top-right of each panel we also write the average
         $1\sigma$ scatter in $\log\,f_j$ at a fixed halo mass for the two galaxy types, also visible
         as shaded areas.}
\label{fig:fj}
\end{figure*}

\begin{figure*}
\includegraphics[width=0.329\textwidth]{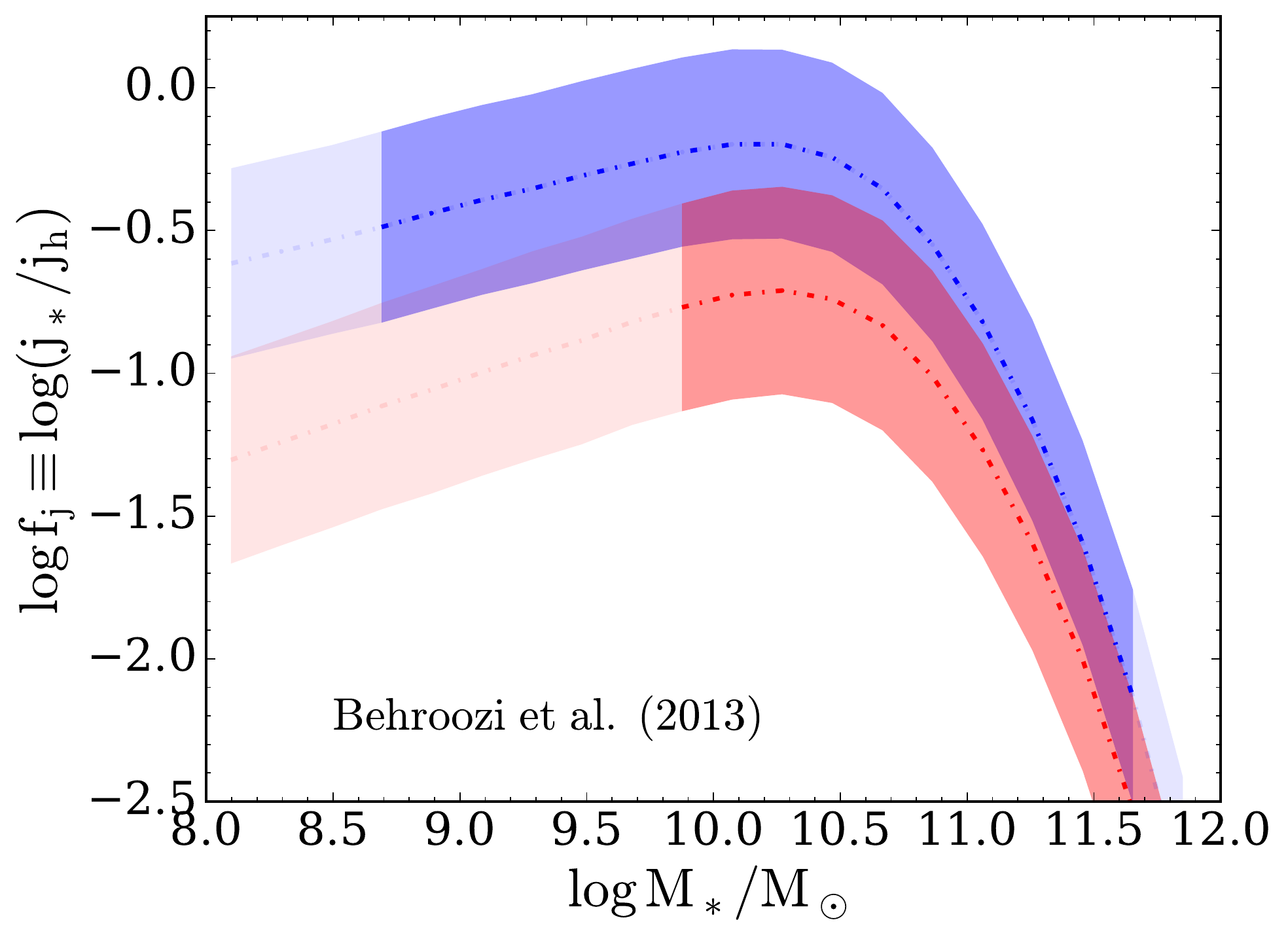}
\includegraphics[width=0.329\textwidth]{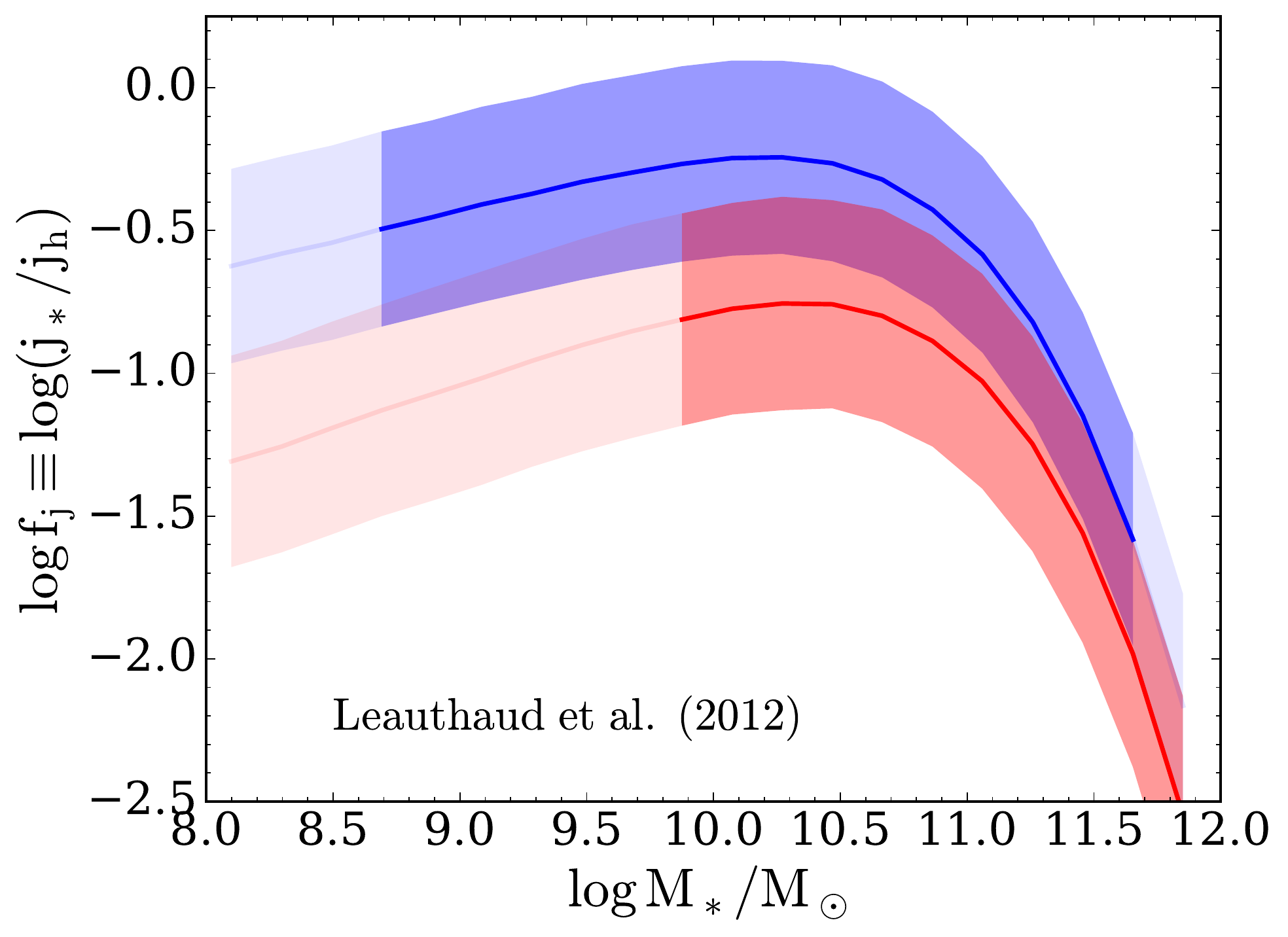}
\includegraphics[width=0.329\textwidth]{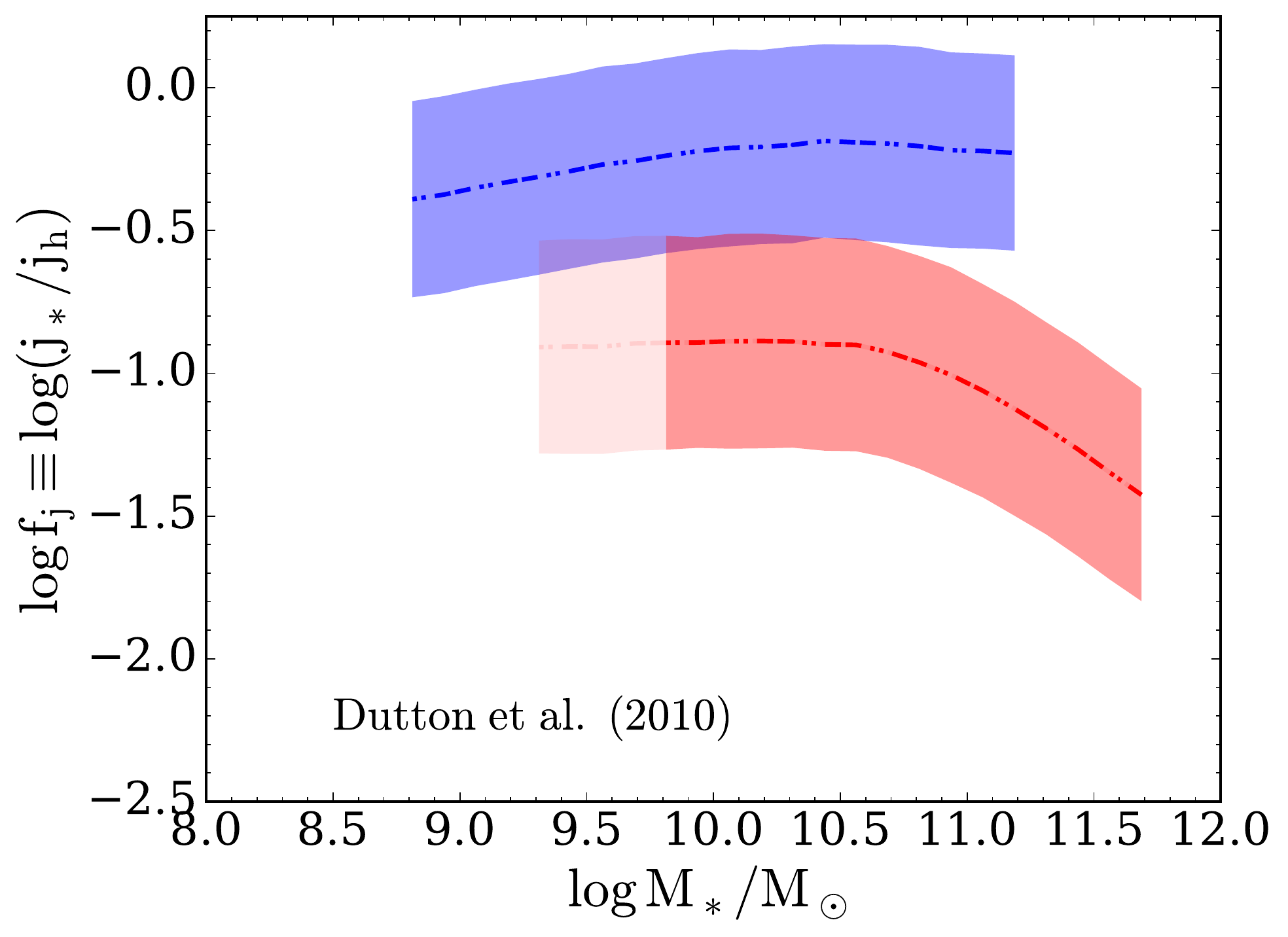} \\
\includegraphics[width=0.329\textwidth]{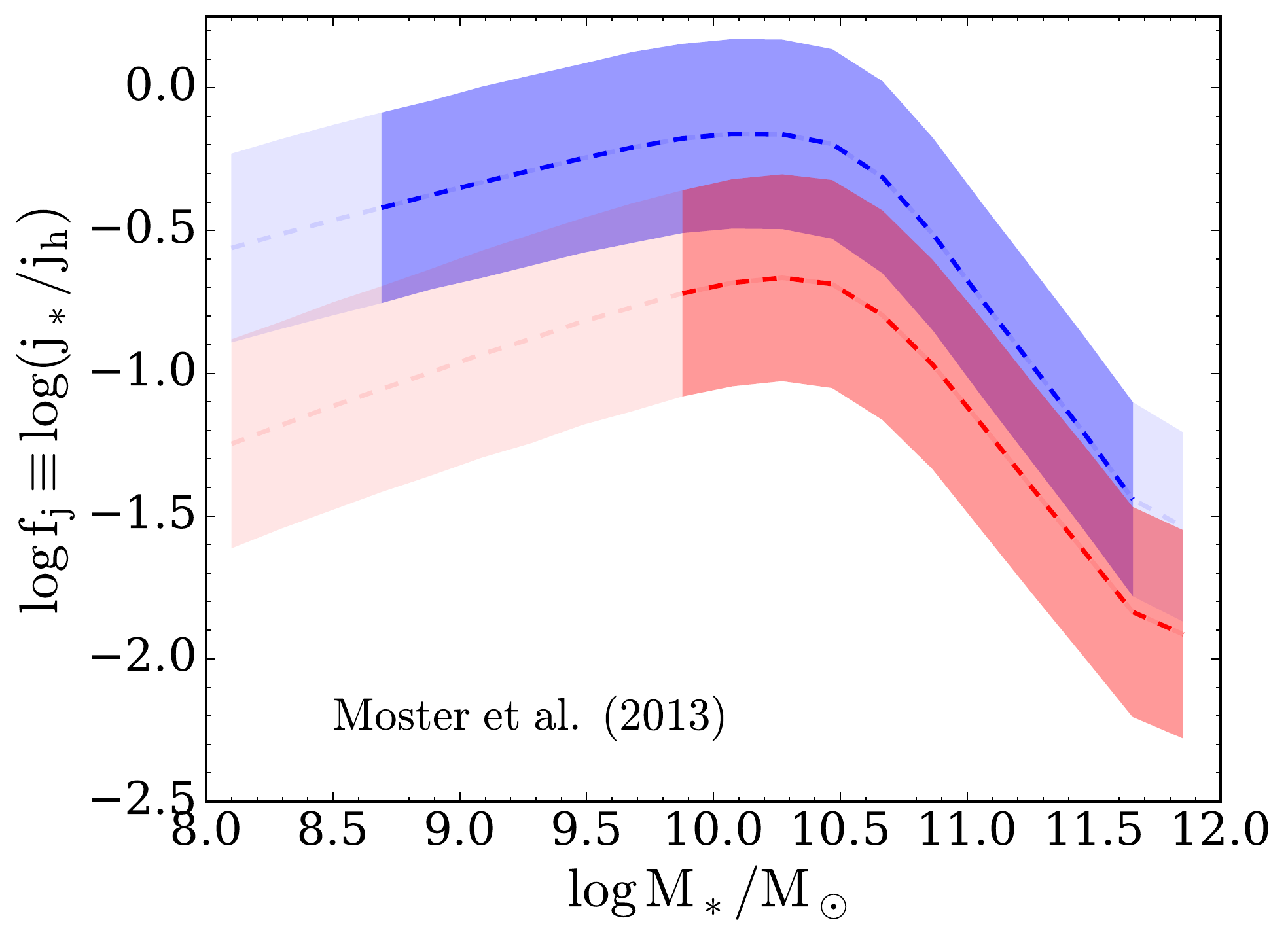}
\includegraphics[width=0.329\textwidth]{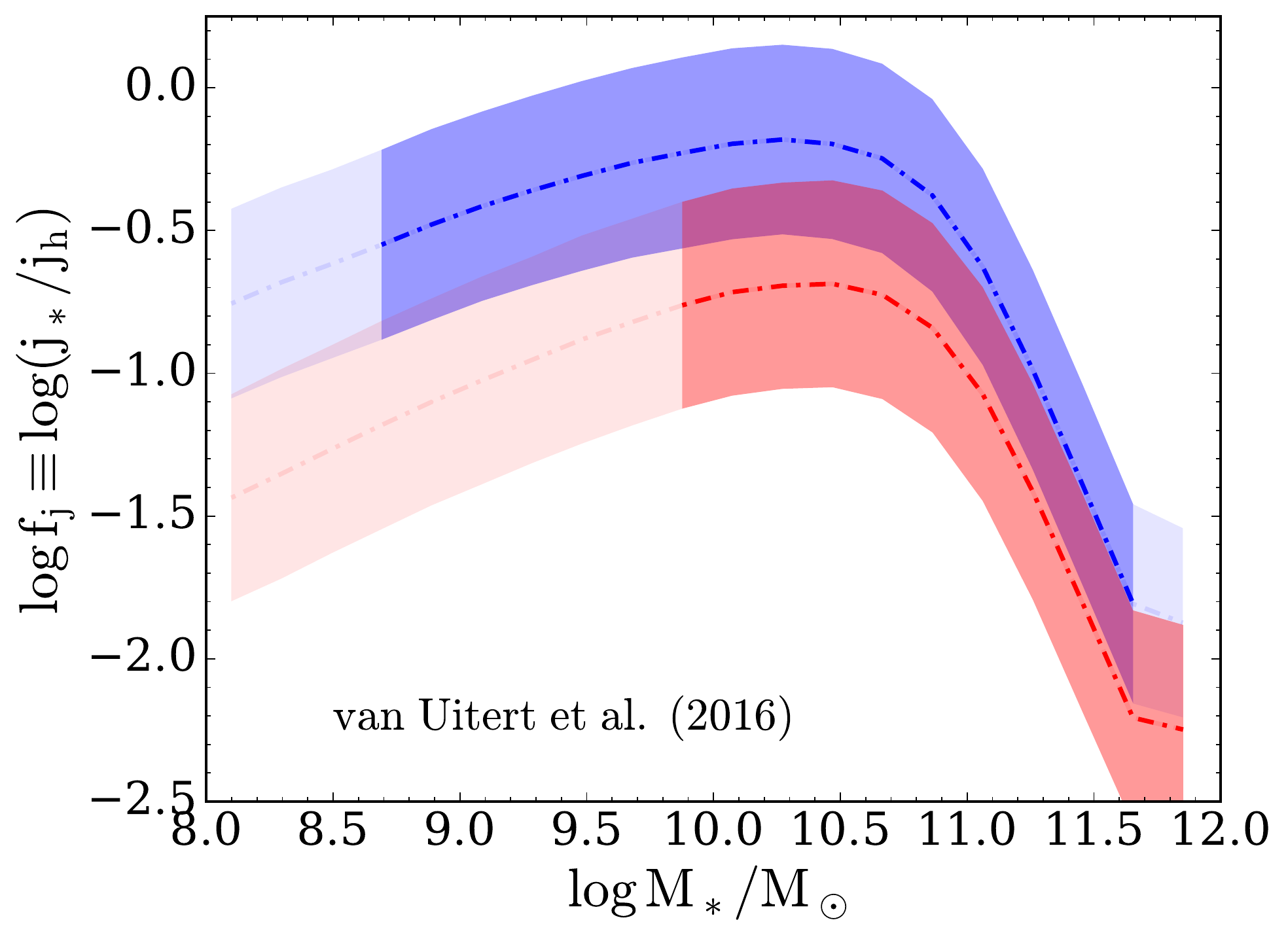}
\includegraphics[width=0.329\textwidth]{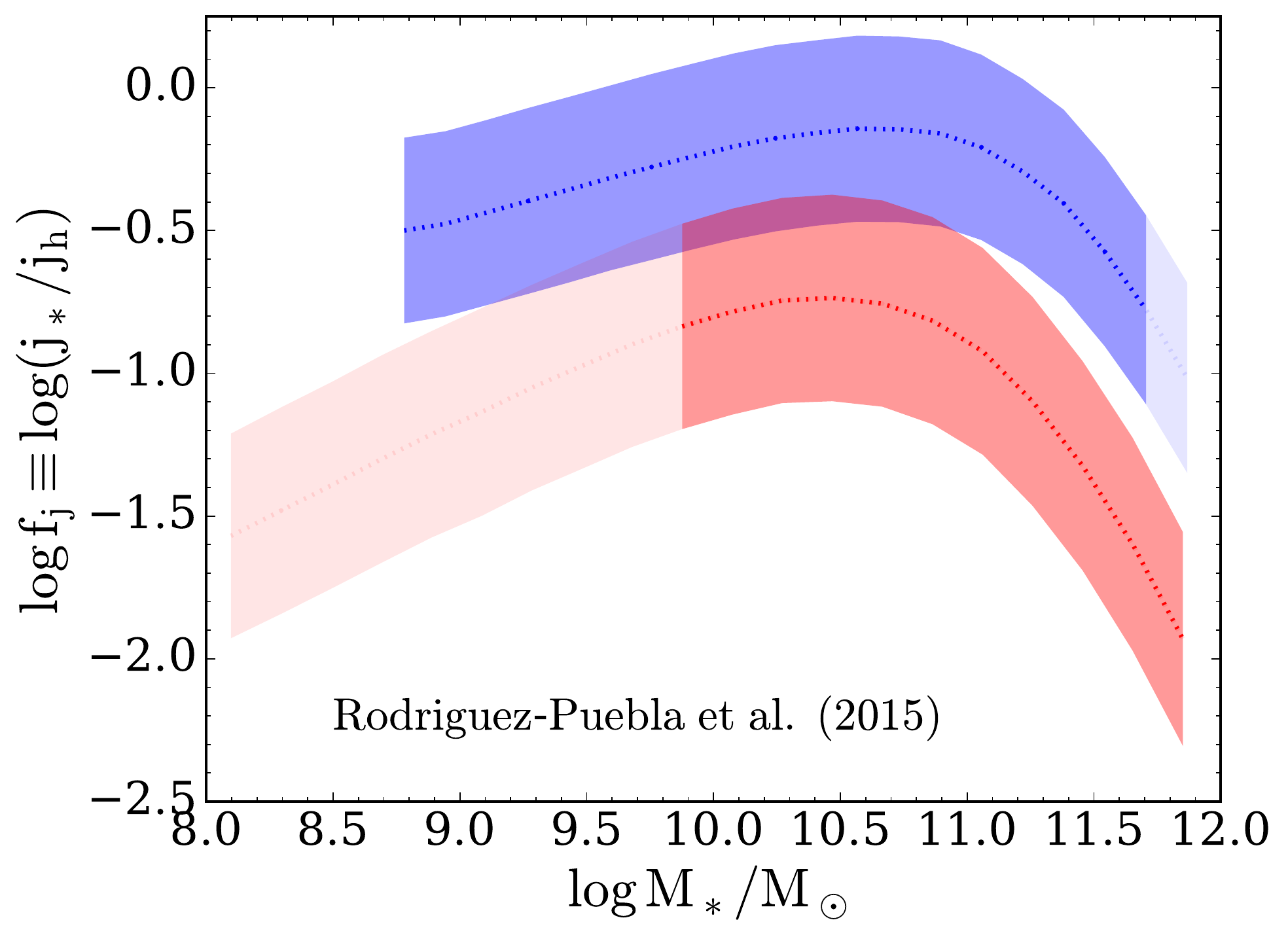}
\caption{Same as Figure \ref{fig:fj}, but as a function of the stellar mass.}
\label{fig:fj-Mstar_RF12}
\end{figure*}

\subsection{The SHSAMR and the distribution of $f_j$ as a function of mass}
\label{sec:fjMh}

Figure~\ref{fig:shsamr} shows the mean relation between the specific angular momentum
of stars and dark matter (SHSAMR) that we obtain assuming six different parametrisations
of the SHMR. This relation is computed separately for spiral and elliptical galaxies and
in both cases it is found to be highly non-linear (in $\log\,\jh-\log\,j_\ast$), irrespectively
of the SHMR adopted. Similarly to the relation between stellar mass and halo mass,
galaxies on average follow a rather steep SHSAMR at low-$\jh$, which becomes shallower for
large galaxies found at higher specific angular momenta.

We now study the full distributions of the retained fractions $f_j$ as a function of halo
mass $\Mh$ (Figure~\ref{fig:fj}) and stellar mass $M_\ast$ (Figure~\ref{fig:fj-Mstar_RF12})
for the six different SHMR. For each SHMR we plot the mean relation for late- and early-type
galaxies together with a shaded area indicating its uncertainty given the current data,
which results from the composition of the scatters in the observed $j_\ast-M_\ast$, the
SHMR, the $c-\Mh$ and the distribution of $\lambda$. We discuss this point in detail in
Section \ref{sec:scatter}.

\begin{table*}
\caption{Best parameters and corresponding $1-\sigma$ uncertainties of a double power-law model
of the type $f_j = \fjzero\left(\Mh/M_0\right)^a (1+\Mh/M_0)^{b-a}$ (equation \ref{eq:fits}) fitted
to the $f_j-\Mh$ relation of the various models in Figure \ref{fig:fj}.}
\label{tab:fits}
\renewcommand{\arraystretch}{1.2}
\begin{tabular}{lcccc}
\hline
Model & $\fjzero$ & $\log M_0/\Msun$ & $a$ & $b$ \\
\hline\hline
\emph{Late type galaxies} & & & & \\
\cite{Behroozi+2013} & $1.9^{+0.05}_{-0.06}$ & $11.65^{+0.06}_{-0.06}$ & $0.67^{+0.05}_{-0.07}$ & $-0.57^{+0.04}_{-0.04}$ \\
\cite{Moster+2013} & $1.96^{+0.05}_{-0.05}$ & $11.54^{+0.05}_{-0.05}$ & $0.6^{+0.08}_{-0.05}$ & $-0.54^{+0.05}_{-0.05}$ \\
\cite{Leauthaud+2012} & $1.35^{+0.04}_{-0.04}$ & $11.85^{+0.06}_{-0.06}$ & $0.44^{+0.03}_{-0.03}$ & $-0.52^{+0.03}_{-0.03}$ \\
\cite{vanUitert+16} & $2.42^{+0.1}_{-0.08}$ & $11.53^{+0.06}_{-0.06}$ & $1.1^{+0.1}_{-0.1}$ & $-0.59^{+0.03}_{-0.03}$ \\
\cite{Dutton+2010} & $1.17^{+0.07}_{-0.06}$ & $11.8^{+0.2}_{-0.2}$ & $0.34^{+0.04}_{-0.05}$ & $-0.19^{+0.05}_{-0.08}$ \\
\cite{RodriguezPuebla+2015} & $2.31^{+0.06}_{-0.06}$ & $11.77^{+0.05}_{-0.05}$ & $0.87^{+0.05}_{-0.06}$ & $-0.52^{+0.02}_{-0.02}$ \\
 & & & & \\
\hline
\emph{Early type galaxies} & & & & \\
\cite{Behroozi+2013} & $0.68^{+0.08}_{-0.06}$ & $11.6^{+0.1}_{-0.2}$ & $0.9^{+0.2}_{-0.2}$ & $-0.6^{+0.1}_{-0.1}$ \\
\cite{Moster+2013} & $0.7^{+0.05}_{-0.06}$ & $11.5^{+0.1}_{-0.1}$ & $0.8^{+0.1}_{-0.2}$ & $-0.52^{+0.05}_{-0.05}$ \\
\cite{Leauthaud+2012} & $0.5^{+0.04}_{-0.04}$ & $11.8^{+0.2}_{-0.2}$ & $0.6^{+0.1}_{-0.1}$ & $-0.5^{+0.1}_{-0.1}$ \\
\cite{vanUitert+16} & $0.92^{+0.1}_{-0.08}$ & $11.5^{+0.2}_{-0.1}$ & $1.5^{+0.4}_{-0.3}$ & $-0.57^{+0.06}_{-0.07}$ \\
\cite{Dutton+2010} & $0.24^{+0.04}_{-0.03}$ & $12.3^{+0.4}_{-0.5}$ & $0.09^{+0.1}_{-0.07}$ & $-0.36^{+0.08}_{-0.1}$ \\
\cite{RodriguezPuebla+2015} & $0.9^{+0.1}_{-0.1}$ & $11.3^{+0.1}_{-0.2}$ & $2.6^{+0.6}_{-0.8}$ & $-0.47^{+0.05}_{-0.05}$ \\
\end{tabular}
\vspace*{15pt}
\end{table*}

At a fixed halo mass, the retained fraction of angular momentum is dependent on galaxy's morphology,
with late-type galaxies having $f_j$ systematically larger than that
of early-type galaxies by roughly a factor $3-5$, consistent with the findings of
\citetalias{RF12}.
For a fixed morphological type, $f_j$ generally has a maximum at the halo mass $\Mstar$
corresponding to the peak of $f_\ast$ (at about $\log\Mh/\Msun\simeq 12$ for all the SHMR
adopted here, see Figure \ref{fig:shmr}). For all the six SHMR the mean $f_j$ tends to mildly
decrease at masses smaller than $\Mstar$, by roughly a factor of $\sim 2$ at $\log (\Mh/\Msun)
= 11$. At masses much larger than $\Mstar$ all models
\citep[but that with the SHMR by][which does not extend to that mass range]{Dutton+2010}
exhibit a sharp decrease of the mean $f_j$, by roughly a factor $3-5$ at
$\log (\Mh/\Msun) = 13.5$. Given the scatter that we estimate for these distributions
(see Section \ref{sec:scatter} for details) the variation of $f_j$ with halo mass is significant
in all models, except those with the SHMR by \cite{Dutton+2010} which are consistent
with a constant $f_j\simeq 0.5$ for spirals and $f_j\simeq 0.1$ for ellipticals. However,
by comparing the top right panel of Figure \ref{fig:fj} with the other panels it is clear that
this result is mainly driven by the small mass range probed by that specific SHMR. From this we
conclude that the finding of \cite{DuttonvandenBosch2012} that $f_j$ is approximately constant with
galaxy mass is mainly due to their choice of the SHMR. Finally, we notice that in the mass range
observationally probed by the Fall relation of \citetalias{RF12} (darker shaded areas and curves
in Figures \ref{fig:fj}-\ref{fig:fj-Mstar_RF12}) also the distribution of
$f_j$ for spiral galaxies with the SHMR of \cite{RodriguezPuebla+2015} appears to be marginally
consistent with a constant $f_j\simeq 0.4$ even if the mean $f_j$ sharply decreases at large
masses. This is because the scatter we have conservatively estimated is rather large (see Section
\ref{sec:scatter}). Similar trends are present in the distribution of $f_j$ as a function of
stellar mass $M_\ast$, thus similar conclusions can be drawn from Figure \ref{fig:fj-Mstar_RF12}.

One of the main results of our work is that the detailed shape of the distribution of $f_j$
as a function of halo mass significantly depends on the SHMR. As discussed in Section
\ref{sec:jstar}, if the Fall relation is a power-law for each morphological type, then $f_j$
\emph{must} decrease for masses larger (and smaller) than $\Mstar$ by an amount that balances
the decrease of $f_\ast^{2/3}$. In particular, $j_\ast\propto M_\ast^\alpha$ and a mass-independent
spin parameter
$\lambda$ for the dark matter yield
\begin{equation}
\label{eq:fj-fstar}
f_j\propto\dfrac{M_\ast^\alpha}{\Mh^{2/3}}=f_\ast^{2/3}\,M_\ast^{\alpha-2/3}.
\end{equation}
In particular, if $f_\ast$ varies as a power-law $f_\ast\propto M_\ast^\beta$ in a given stellar
mass range, then in the same range $f_j$ also behaves as a power-law
\begin{equation}
\label{eq:fj-Mstar}
f_j \propto M_\ast^{\alpha + \frac{2}{3}(\beta-1)},
\end{equation}
or, equivalently, if $f_\ast$ varies as $f_\ast\propto \Mh^\gamma$ in a given halo mass range, then
$f_j$ goes as
\begin{equation}
\label{eq:fj-Mh}
f_j \propto \Mh^{\alpha(\gamma+1) - \frac{2}{3}}.
\end{equation}
In the stellar mass range probed by \citetalias{RF12} the Fall relations for spirals and ellipticals
have $\alpha=0.52$ and $\alpha=0.60$, respectively. For spiral galaxies, for instance, this implies
that for a \cite{Dutton+2010} SHMR the retained fraction $f_j$ roughly behaves as $f_j\propto
\Mh^{0.37}$ and $f_j\propto \Mh^{-0.15}$ at low- and high-masses respectively; while for a
\cite{Moster+2013} SHMR equations \eqref{eq:fj-Mstar}-\eqref{eq:fj-Mh} imply the behaviours
$f_j\propto\Mh^{0.57}$ and $f_j\propto\Mh^{-0.46}$ at low- and high-masses respectively.

Our results are a consequence of the fact that the empirical Fall relation is well represented
by a single power-law of index close to $2/3$. All the observations available at the moment are
consistent with a power-law $j_\ast-M_\ast$ relation, whose normalization (and maybe logarithmic
slope) vary with galaxy morphology. However, it is not definitively clear whether an upward
bend of such relation for very massive ($\log M_\ast/\Msun>11.5$) discs can be completely ruled
out \citep[see e.g.][]{FallRomanowsky2013} and the determination of $j_\ast$ for the most massive
nearby disc galaxies known would be extremely helpful in shedding some light on this issue.
At stellar masses smaller than $\sim 10^9\Msun$ observational constraints on both the Fall relation
and the SHMR are either very uncertain or missing, hence our conclusions can not reliably extend much
further down this mass. However, if we were to extrapolate the behaviour of both the Fall relation and
the SHMR to the dwarf galaxies regime, we would find that $f_j$ decreases with decreasing galaxy mass.
At the moment there are only some indications that the Fall relation may be extended through the dwarf
regime \citep[][]{ChowdhuryChengular2017} and some studies which estimate the SHMR for dwarfs
and compare it to that of more massive galaxies \citep[e.g.][]{Miller+2014}, so we are not able to
draw any significant conclusion here. However, we notice that our expectation of a decreasing
retained fraction $f_j$ with decreasing stellar mass below $10^9\Msun$ is in agreement with the
recent hydrodynamical simulations of \cite{ElBadry+2017}, whose model galaxies closely follow the
SHMR of \citet{Moster+2013}, at least for $M_\ast> 10^{8.5}\Msun$.

\subsubsection{A simple description of the $f_j-\Mh$ relation}
\label{sec:fits}

We find that a reasonable description of the mean $f_j$ as a function of $\Mh$ is provided by
the following double power-law model
\begin{equation}
\label{eq:fits}
f_j = \fjzero \left(\frac{\Mh}{M_0}\right)^a \left(1+\frac{\Mh}{M_0}\right)^{b-a}
\end{equation}
where $\fjzero$, $M_0$, $a$ and $b$ are constants representing respectively a normalization,
a mass scale and the low- and high-mass slopes. In Table \ref{tab:fits} we report the best-fit
values of the above parameters to the twelve distributions of $f_j$ as a function of halo mass,
in the range of stellar masses probed by the \citetalias{RF12} Fall relations, presented in
Figure \ref{fig:fj} (found by minimising a $\chi^2$ likelihood) together with their $1\sigma$
uncertainties (computed by sampling the posterior distribution using a Monte Carlo
Markov Chain method, assuming uninformative priors). The values of the best-fitting low-mass
and high-mass slopes $a$ and $b$ can be compared to the scaling given in equation
\eqref{eq:fj-Mh} for a power-law Fall relation and in the regimes where the SHMR is approximated
by a power-law.

\subsection{Scatter of the SHSAMR}
\label{sec:scatter}

The large shaded areas in Figure \ref{fig:fj} derive from a combination of the intrinsic
and observational scatters in the $j_\ast-M_\ast$ plane and SHMR, as well as from the intrinsic
scatter of the theoretical $c(\Mh)$ relation and $\lambda$ distribution. Each of these scatters
sum up in quadrature to yield the final width
\begin{equation}
\label{eq:scatter1}
{\tilde{\sigma}_{f_j}} = \sqrt{\sigma_{\rm j_\ast, obs.}^2+\sigma_{\rm SHMR}^2+
\sigma_{c-\Mh}^2+\sigma_\lambda^2} \sim 0.32-0.36\mbox{  dex,}
\end{equation}
While encompassing the current uncertainties, this should not be considered as an
estimate of the intrinsic scatter of the SHSAMR $\sigmafj$, unless one makes the assumption
that all the terms in equation \eqref{eq:scatter1} are uncorrelated, which is not very realistic,
considering that at least the observed scatter $\sigma_{\rm j_\ast, obs.}$ will most likely
depend on $\sigma_{\rm SHMR}$, $\sigma_{c-\Mh}$ and $\sigma_\lambda$. 
Thus, the width of the shaded areas should be regarded as a very conservative
\emph{upper limit} on the intrinsic scatter of the $f_j-\Mh$ relation.

The observed scatter $\sigma_{\rm j_\ast, obs.}$ of the $j_\ast-M_\ast$ relation is found to be
much smaller than the sum in quadrature of $\sigma_{\rm SHMR}$, $\sigma_{c-\Mh}$ and
$\sigma_\lambda$ (see Table \ref{tab:scatter}). This suggests that the scatters of these
quantities are likely to be correlated in such a way that when combined as in equation
\eqref{jstarMstar_theory} they yield a scatter of $\sim 0.19$ dex for the $j_\ast$ of spirals
and $\sim 0.24$ dex for that of ellipticals. 
Note that these conclusions are mostly driven by the scatter in $\lambda$ (see Table
\ref{tab:scatter}), which is theoretically well established.

\begin{figure*}
\includegraphics[width=0.329\textwidth]{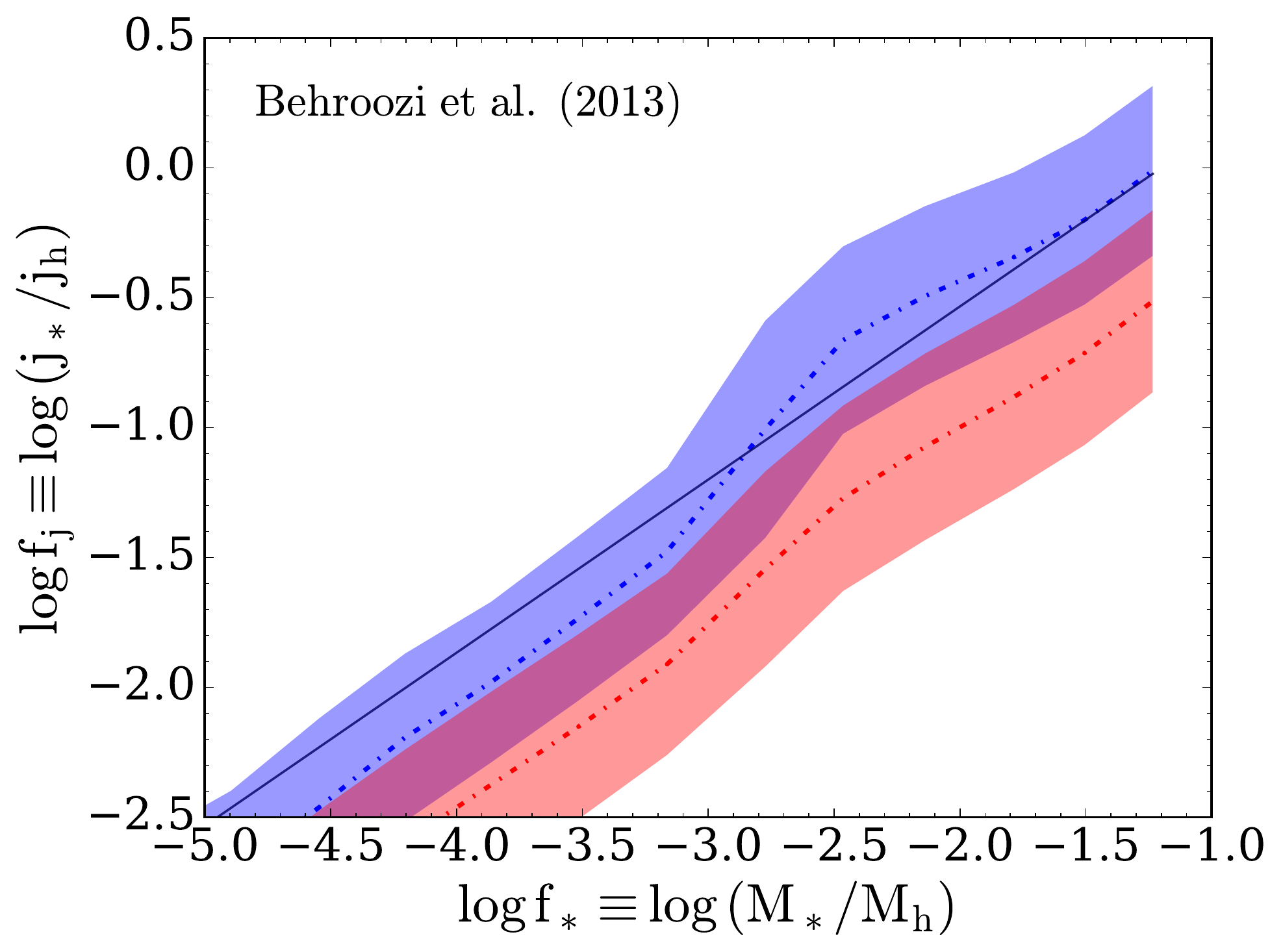}
\includegraphics[width=0.329\textwidth]{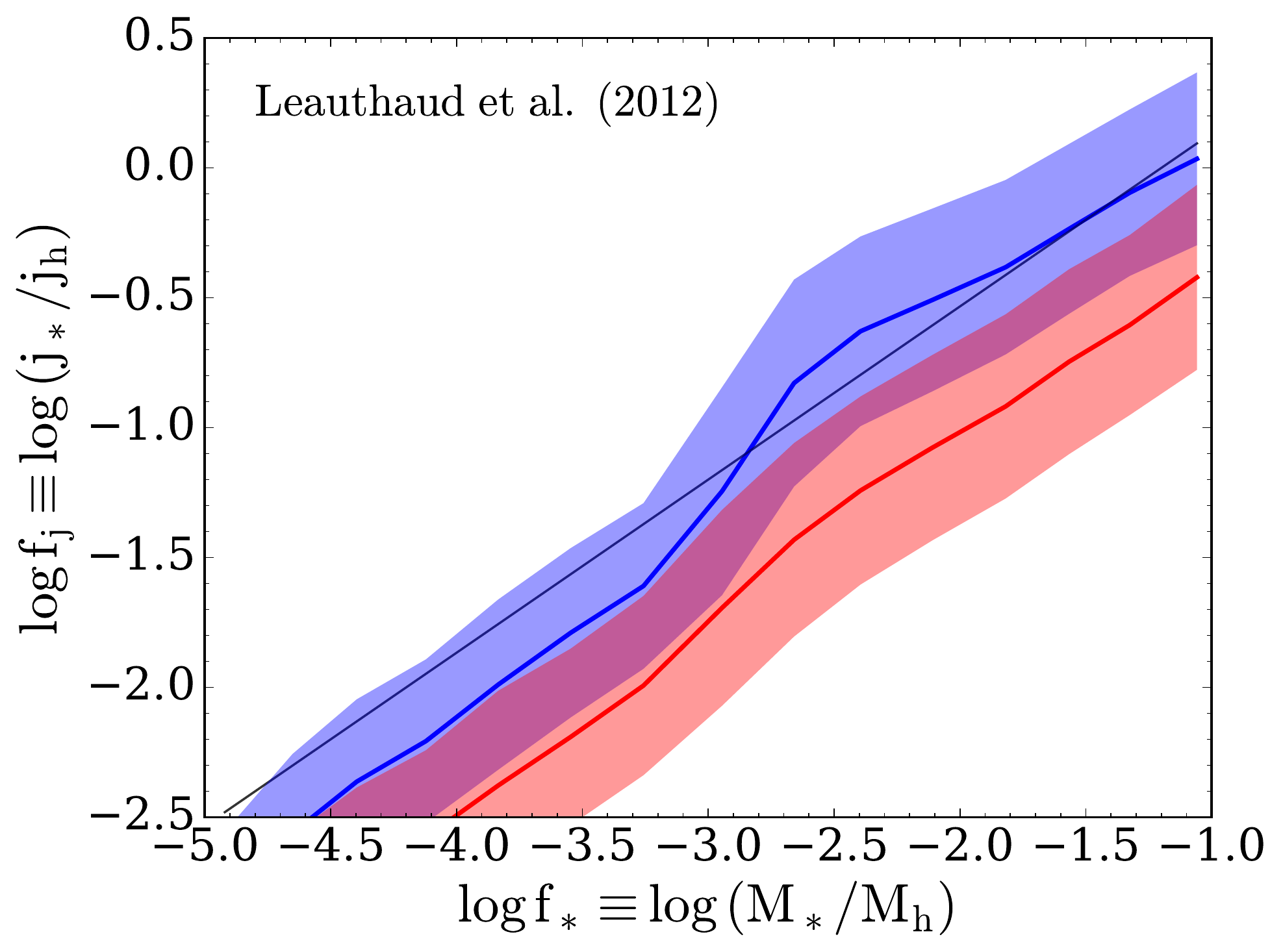}
\includegraphics[width=0.329\textwidth]{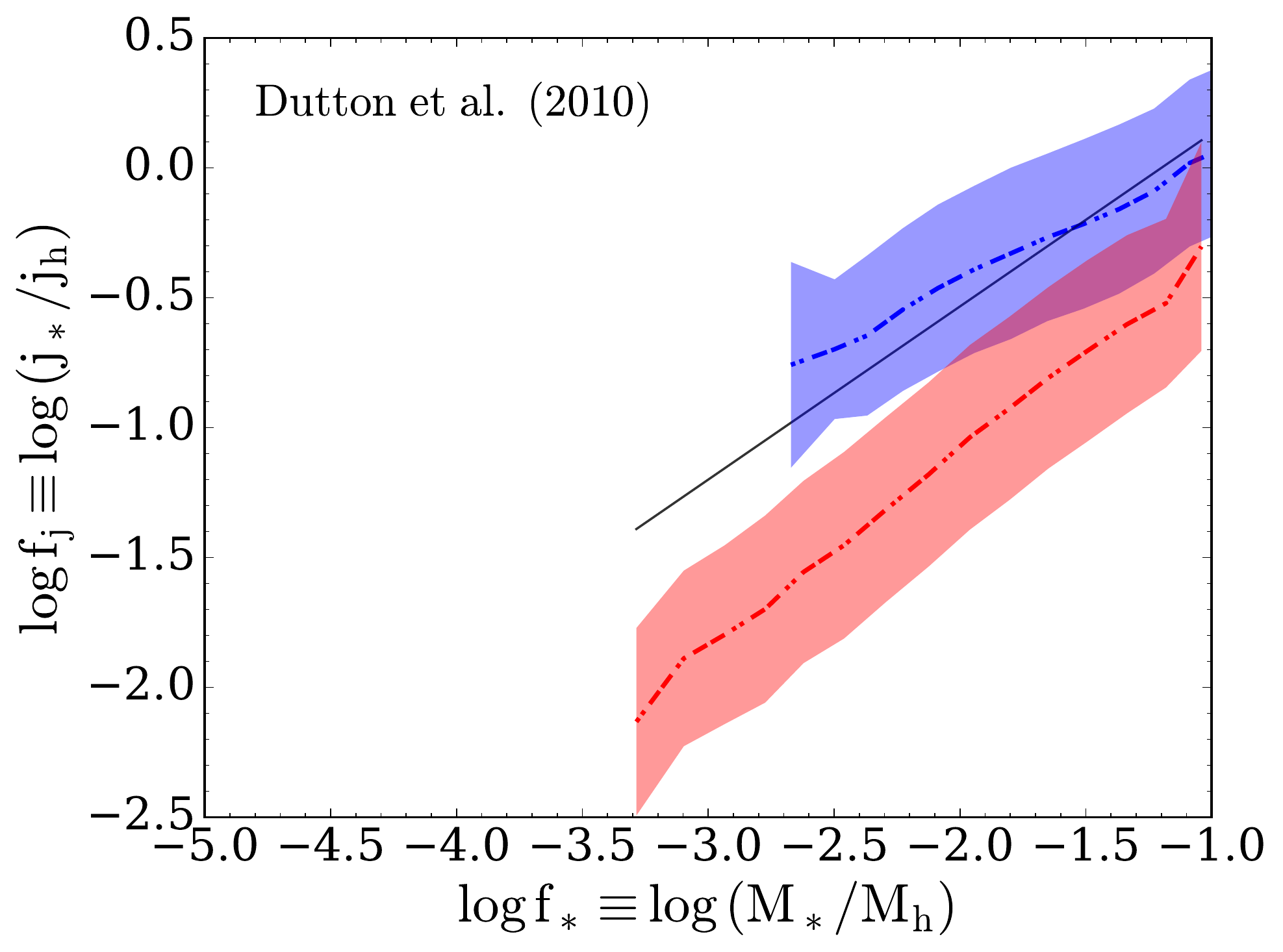} \\
\includegraphics[width=0.329\textwidth]{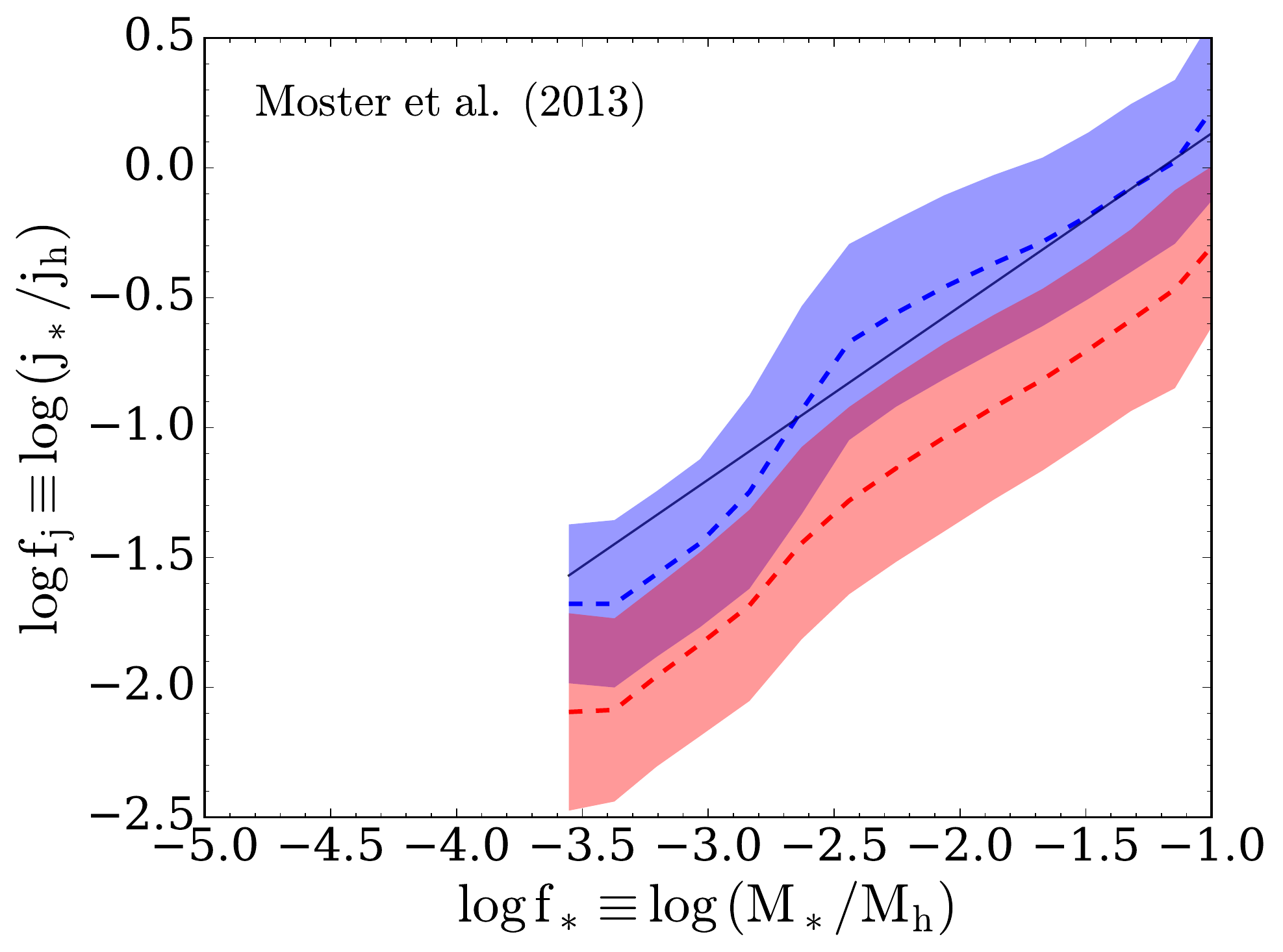}
\includegraphics[width=0.329\textwidth]{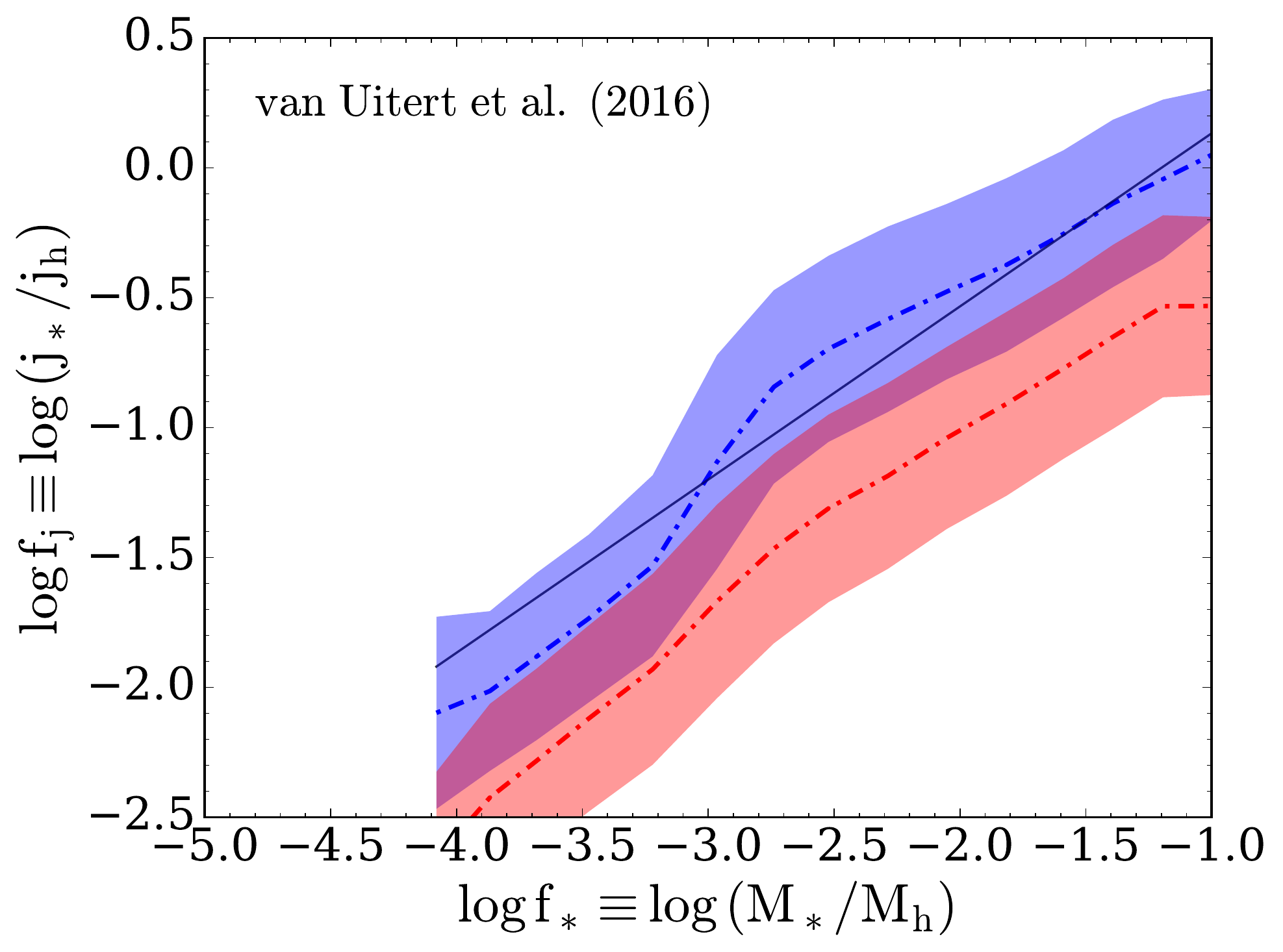}
\includegraphics[width=0.329\textwidth]{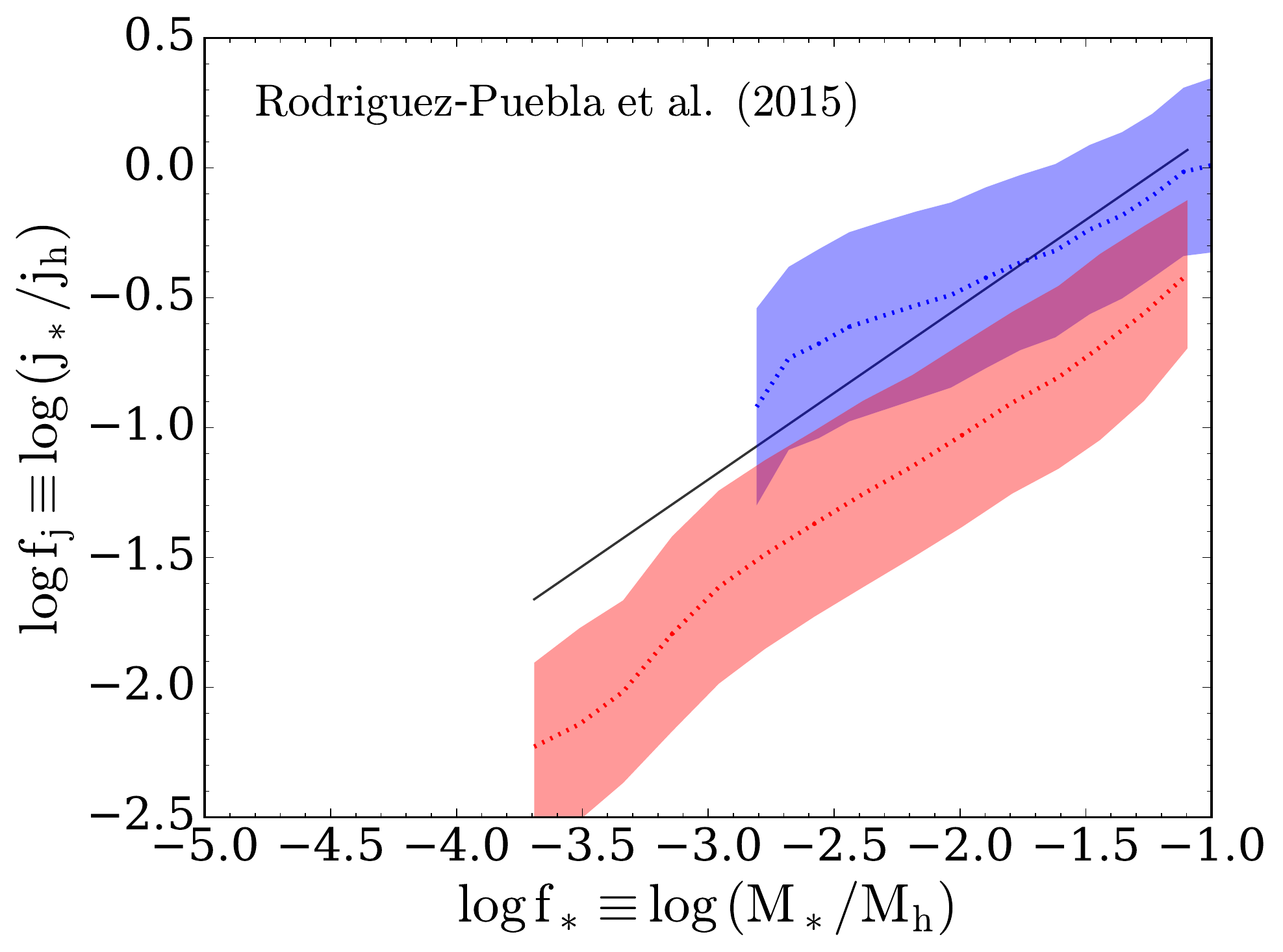}
\caption{Retained fraction of angular momentum as a function of the stellar fraction, or star
		 formation efficiency, for the same models as in Figure
         \ref{fig:fj}. We also show the proportionality $f_j\propto f_\ast^{2/3}$
         as a grey thin solid line.}
\label{fig:fj-fstar_RF12}
\end{figure*}

\begin{table}
\begin{center}
\caption{Scatter budget (in dex) as in equations \eqref{eq:scatter1}-\eqref{eq:scatter2}.
The scatter of the SHMR are for the model of \citet{RodriguezPuebla+2015} and observational
estimates are from \citetalias{RF12} (see equations \ref{eq:jstar-Mstar_rf12_ltg} and
\ref{eq:jstar-Mstar_rf12_etg} above).}
\label{tab:scatter}
\begin{tabular}{lcccc}
\hline
Galaxies & $\sigma_\lambda$ & $\sigma_{c-\Mh}$ & $\sigma_{\rm SHMR}$ & $\sigma_{\rm j_\ast, obs.}$ \\
\hline
\emph{Late types} & 0.25 & 0.11 & 0.11 & 0.19 \\
\emph{Early types} & 0.25 & 0.11 & 0.14 & 0.24 \\
\end{tabular}
\vspace*{15pt}
\end{center}
\end{table}

The considerations above are critical also to interpret the results on the mass-dependence of
the retained fraction. For instance, the $f_j$ distribution of spirals in the case of the SHMR
of \cite{RodriguezPuebla+2015} is consistent with a constant only if the intrinsic scatter of
the SHSAMR is as large as the displayed shaded area, which we argued above to be rather unlikely.

\subsection{The distribution of $f_j$ as a function of $f_\ast$}
\label{sec:fjfstar}

In Figure \ref{fig:fj-fstar_RF12} we show the retained fraction of angular momentum $f_j$ as a
function of the stellar fraction, or star formation efficiency, $f_\ast$. These two quantities
are clearly correlated such that galaxies with smaller stellar fraction
also have smaller retained fraction of angular momentum.
The correlation roughly follows $f_j\propto f_\ast^{2/3}$, so that the quantity $Q$ defined
in equation \eqref{eq:const_fjfstar} is about constant with galaxy mass. The deviations of
the observed Fall relation from $j_\ast\propto M_\ast^{2/3}$ can be interpreted as a consequence
of the deviations of the $f_j-f_\ast$ relation from $f_j\propto f_\ast^{2/3}$ (see equation
\ref{eq:fj-fstar}).

Note that similar conclusions to ours can be reached by modelling galaxies that are on the
mass-rotational velocity relation, i.e. the \cite{TF77} relation, within a $\Lambda$CDM framework.
Combining a $M_\ast\propto \Vrot^3$ relation with the scalings for dark matter haloes
summarized in Section \ref{sec:lcdm} one finds $f_j\propto f_v^2$ where $f_v\equiv\Vrot/\Vmax$,
which also immediately yields $f_j\propto f_\ast^{2/3}$ \cite[see][]{NavarroSteinmetz2000}.
Currently this type of models are able to reasonably represent the observed Tully-Fisher,
mass-size and SHMR simultaneously \citep[see][]{Ferrero+2017}. However, even if the Tully-Fisher
relation is much tighter and much better known w.r.t. the Fall relation, unlike the angular momentum
retention fraction $f_j$ which is directly related to the main phenomena driving galaxy evolution
(e.g. the accretion history and feedback), the parameter $f_v$ has no clear physical meaning
and its link with galaxy evolution is much more obscure.

\begin{figure*}
\includegraphics[width=0.49\textwidth]{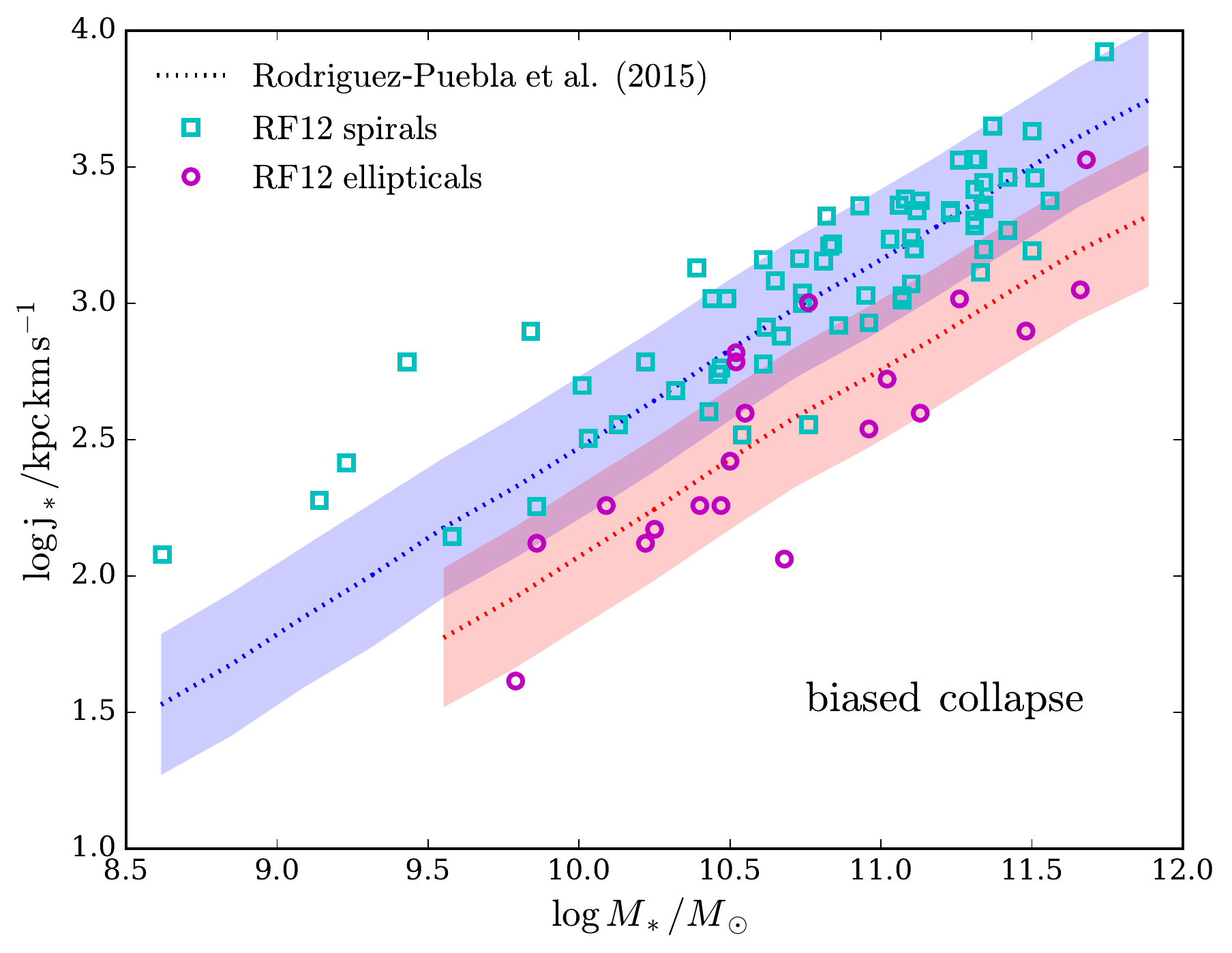}
\includegraphics[width=0.49\textwidth]{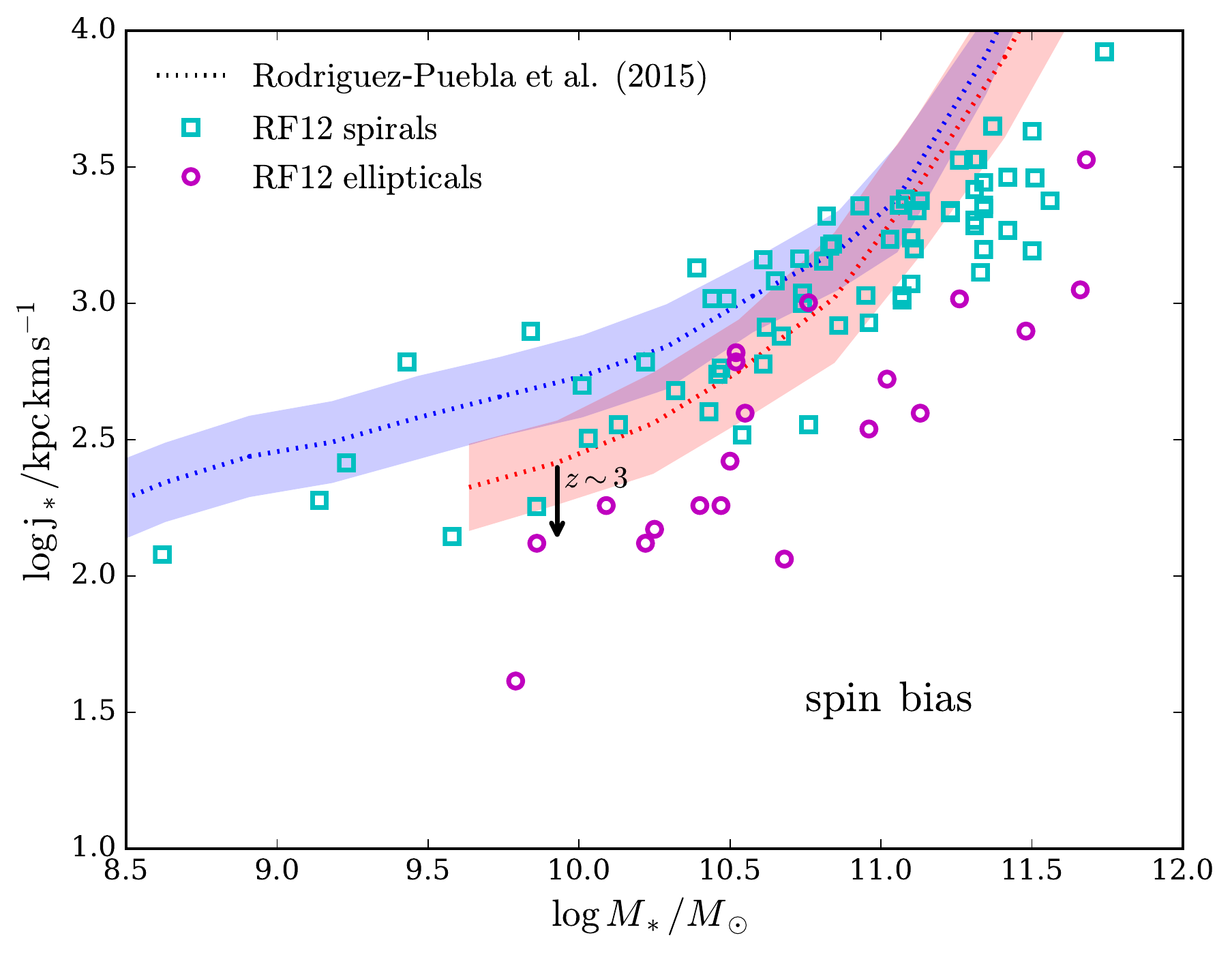}
\caption{\emph{Left-hand panel:} Fall relation for
		 the biased collapse model in which $f_j\propto f_\ast^{s}$ (see Section
         \ref{sec:frwd_fjfstar}). The blue and red shaded areas are the 1$\sigma$ predictions
         of the models (the dotted curve is the mean), while the cyan squares and magenta
         circles are the late- and early-type galaxies observed by \citetalias{RF12}. The
         models employ the SHMR of \citet{RodriguezPuebla+2015} for each galaxy type.
         \emph{Right-hand panel:} same as the left-hand panel, but for the spin-bias
         model in which the halo spin correlates with galaxy type (see Section
         \ref{sec:frwd_lambda}). In this model we adopt a common $fj = 0.5$ for spirals and
         ellipticals, while the black arrow shows the effect of assuming that all ellipticals were
         already quenched at $z = 3$.}
\label{fig:forward_models}
\end{figure*}

The $f_j-f_\ast$ relation indicates that \emph{galaxies which are globally efficient at forming stars
are also efficient at ``retaining'' angular momentum}. Hence the formation history of a galaxy which
has very efficiently turned its available gas into stars is such that its spin is comparable
to that of the dark halo (and viceversa), as cosmological hydrodynamical simulations also suggest.
Recent studies have found that such galaxy/halo mass- and spin-connection is regulated either by the
effect of stellar feedback, which preferentially removes the low-$j$ material \citep[see][]
{PedrosaTissera2015,DeFelippis+2017}, together with the halo assembly history, since mergers
tend to lower the angular momentum of stars \citep[see][]{Sokolowska+2017}, and/or by the accretion
history of gas onto the galaxy, which proceeds from the low-$j$ material first and then continues
to the high-$j$ one for systems which are still efficient at forming stars at later times
\citep[see][]{ElBadry+2017}. In Section \ref{sec:forward_models} we expand the discussion of this
latter kind of models, which we call \emph{biased collapse} models following \citetalias{RF12}.

All these factors potentially play key roles in shaping galaxies as we see them today and they
must be considered when attempting to explain the current morphology and structural properties
of galaxies. Any galaxy formation model, which aims at explaining the formation and evolution
of galaxies in terms of their history of accretion and star formation, has necessarily to deal
with the fact that the global star formation efficiency is intimately linked to the global
efficiency at angular momentum retention.

\section{Physical models for the angular momentum of galaxies}
\label{sec:forward_models}

In this Section we describe two simple physically-based models for the angular momentum content
of galaxies and we compare their predictions on the specific angular momentum-mass diagram
with observations from \citetalias{RF12}. This approach is complementary to that adopted in
Section \ref{sec:ret_frac} and may be useful to test the predictions of some specific scenarios.
In particular, we first ask whether a power-law Fall relation naturally emerges in a
\emph{biased collapse scenario}, in which the global efficiency of star formation correlates
with the efficiency at angular momentum retention (Section \ref{sec:frwd_fjfstar}); then we
ask whether the different normalization of the Fall relations for spirals and ellipticals can
be ascribed to a \emph{spin-bias} scenario and not to a variable $f_j$ with galaxy mass
(Section \ref{sec:frwd_lambda}).
For simplicity, here we use only the SHMR of \cite{RodriguezPuebla+2015} and we are
implicitly assuming that their blue and red galaxy populations trace the spirals and
ellipticals of \citetalias{RF12}.

\subsection{Biased collapse models: a power-law $f_j - f_\ast$ relation}
\label{sec:frwd_fjfstar}

We first test a simple model in which $f_j$ varies with galaxy mass.
$f_j$ and $f_\ast$ are strongly correlated both in an empirical $\Lambda$CDM model based on
the Fall relation (Section \ref{sec:ret_frac} and Figure \ref{fig:fj-fstar_RF12}) and in
one based on the Tully-Fisher relation \citep[see][and Section \ref{sec:fjfstar}]
{NavarroSteinmetz2000}. The same happens also in a model in which the global star formation
efficiency depends on the gas cooling efficiency: if the gas has a cumulative angular
momentum distribution $j_{\rm gas}(<r) \propto M_{\rm gas}(<r)^s$ and it is accreted (and cools)
onto the galaxy inside-out, then \citep[see][]{vandenBosch1998,DuttonvandenBosch2012}
\begin{equation}
\label{eq:biasedcollapse}
f_j = (f_\ast/f_{\rm b})^s,
\end{equation}
where $f_{\rm b} \equiv \Omega_\textrm{b}/\Omega_\textrm{m}$ is the cosmic baryon fraction.
Hence $f_\ast$ is expected to correlate with $f_j$ if stars form preferentially from the gas
at highest densities, i.e. closest to the centre where the gas cooling was more effective
\citep[e.g.][see also \citetalias{RF12}, Section 6.3.2]{Kassin+2012}.
We refer to this scenario as \emph{biased collapse}.

In the model presented in this Section we interpret $f_j$ as the \emph{fraction of the halo's
specific angular momentum present in the gas actually able to cool and form stars in the galaxy}.
In practice, we are considering galaxies formed out of gas at angular momentum $j$
between $0\leq j \leq j_{\rm max} < j_{\rm vir,DM}$, where $j_{\rm max}$ is the maximum
specific angular momentum of the material that collapsed to form the central galaxy. A correlation
between $f_j$ and $f_\ast$ is expected also in the presence of feedback \citep[as it is seen
in cosmological simulations, e.g.][]{PedrosaTissera2015}, though it may alter the value of
the exponent $s$ in equation \eqref{eq:biasedcollapse} due to angular momentum mixing (see below).

Motivated by these results we make the ansatz:
\begin{equation}
\label{eq:frwd_fjfstar}
\log f_j = A + s\log f_\ast,
\end{equation}
We repeat the steps (i) to (iv) as in Section \ref{sec:mod_summary}, with the SHMR of
\cite{RodriguezPuebla+2015}; then, we assign $f_j$ to each galaxy as in equation
\eqref{eq:frwd_fjfstar}, with variable $A$ and $s$ and no scatter;
finally, we use equation \eqref{jstarMstar_theory} to compare with the observations of
\citetalias{RF12}.

We determine $A$ and $s$ by minimising a $\chi^2$ likelihood and we also allow for the
presence of outliers in the observed data. Then, we compute the posterior probabilities
of the two parameters by assuming flat priors (using a Monte Carlo Markov Chain method).
We obtain
\begin{align}
\label{eq:AB_ltg}
A&=0.65 \pm 0.13,& \,\, s&=0.68 \pm 0.05,& \,\, (\mbox{spirals})\\
\label{eq:AB_etg}
A&=0.23 \pm 0.12,& \,\, s&=0.69 \pm 0.05.& \,\, (\mbox{ellipticals})
\end{align}
Our model estimates a fraction of $f_{\rm out} = 0.10\pm 0.06$ of possible outliers in
the Fall relation for spirals at $\log M_\ast\lesssim 10$, while it is consistent with
zero for ellipticals.  The slope of the relation \eqref{eq:frwd_fjfstar} is unsurprisingly
in agreement with the value $2/3$ for both galaxy types: in fact, this is the only value
of $s$ that generates a power-law Fall relation (also with slope $2/3$). For reference,
the case $f_j=1$ would correspond to $s = 0$ in equation \ref{eq:frwd_fjfstar}
\citep[e.g.][\citetalias{RF12}]{MMW98} and we find it to be disfavoured by observations.
On the other hand, our best fit value $s \simeq 0.7$ is significantly smaller than the
value $s\simeq 1.3$ suggested by \citep[][]{DuttonvandenBosch2012}, under the assumption
that the baryons and the dark matter have exactly the same angular momentum distribution.
Note that equation \eqref{eq:biasedcollapse} is consistent with an angular momentum
distribution of baryons $dM/dj \propto j^q$ with $s = 1/(1+q)$. The best fit $s = 0.68$
would imply $q = 0.47$, which is interestingly in between the values $q \simeq 0$ and
$q \simeq 1$ expected for a dark matter halo and a pure disc, respectively \citep[e.g.][]{vdBBS2001}.
This may be an indication of mixing processes, possibly related to feedback, having
redistributed the angular momentum of baryons, resulting in a shallower angular momentum
distrbution \citep[e.g.][]{Brook+2012}.

Remarkably, the value of the normalization for spirals is consistent with $A = -0.68 \times \log(f_b) = 0.55$, as expected from equation \eqref{eq:biasedcollapse} with $s = 0.68$. For ellipticals, dynamical friction induced by mergers may have contributed to net losses of angular momentum resulting in an overall smaller normalisation.

We show in the left-hand panel of Figure \ref{fig:forward_models} the predictions of this
model compared with the galaxies observed by \citetalias{RF12}. A remarkable agreement
can be appreciated. Though, the scatter of the predicted relation is still too large compared to
the observed scatter: the model predicts $0.26$ dex w.r.t. the observed $0.19$ dex and
$0.24$ dex for late and early types.
This is because, as discussed in Section \ref{sec:scatter}, we
are overestimating the scatter of the Fall relation since we are assuming the
scatters in the SHMR, $\lambda$ distribution, $c(\Mh)$ and $f_j-f_\ast$ relations to be
uncorrelated.
Moreover, even with this simplified assumption of no-intrinsic scatter in the $f_j-f_\ast$
relation we are still overpredicting the scatter in the $j_\ast-M_\ast$ relation.

\subsection{Spin-bias models: low-rotating galaxies in low-spin haloes}
\label{sec:frwd_lambda}

Another hypothesis one could envisage is that the retention fraction $f_j$ is actually
constant with mass and galaxy type and, instead, the scatter of the halo spin distribution
correlates with galaxy type and specific angular momentum. In this picture, at a fixed
halo mass, haloes with larger spins host preferentially galaxies of later morphological types
and with larger specific angular momenta \citepalias[see e.g.][Section 6.2]{RF12}.
We refer to this scenario as \emph{spin-bias}.

To test this hypothesis, we consider a model in which, for a given halo mass, the haloes with
the largest $\lambda$ are occupied by spirals and those with the smallest $\lambda$ are occupied
by ellipticals. In our simplified picture of galaxies of two morphological types well traced
by their colours, we can define for each halo mass a critical value $\lambda_{\rm crit}$ which
divides haloes populated by galaxies of the two types.
This must be a function of mass $\lambda_{\rm crit}(\Mh)$ since haloes of different masses have
different probabilities to contain galaxies of a given type \citep[i.e. morphology depends
on galaxy mass, e.g.][]{Kelvin+2014}.
To find the function $\lambda_{\rm crit}=\lambda_{\rm crit}(\Mh)$ we use an iterative procedure.
Starting from a given SHMR and an observed distribution of late/early types fraction as a function
of stellar mass $f_T=f_T(M_\ast)$, taken from \cite{Kelvin+2014}, we proceed as follows:

\begin{itemize}
\setlength\itemsep{0.5em}
\item[(i)] we generate a uniform population of $N_{\rm haloes}=10^5$ in
		   $10\leq \log\Mh/\Msun \leq 14$ and we divide it into 12 equally-spaced bins;
\item[(ii)] starting from a flat guess for the function $\lambda_{\rm crit}(\Mh)$, in a given halo
			mass bin we assign to haloes with $\lambda>\lambda_{\rm crit}(\Mh)$ the type ``spiral''
            and the type ``elliptical'' to haloes with $\lambda<\lambda_{\rm crit}(\Mh)$;
\item[(iii)] we compute stellar masses for each halo using the SHMR of \cite{RodriguezPuebla+2015}
			 for the two galaxy populations;
\item[(iv)] we bin again the entire population, but this time in stellar mass; in each bin we compute
            the fraction of spirals and ellipticals as determined by our guess function
            $\lambda_{\rm crit}(\Mh)$ and we compare the result to the observed distribution from
            \cite{Kelvin+2014};
\item[(v)] we update the guess $\lambda_{\rm crit}(\Mh)$, iterating the steps above, until a
           satisfactory match to the target $f_T(M_\ast)$ of \cite{Kelvin+2014} is found.
\end{itemize}

With this procedure we have defined a model galaxy population for which the morphological type
distribution is matched to the halo spin distribution consistently with the given SHMR. Now we
use equation \eqref{jstarMstar_theory}, with an arbitrary constant $f_j=0.5$ for both galaxy type
and mass, to predict the Fall relation.

We also tried with a different approach, that is, starting from a stellar mass distribution,
in each stellar mass bin we assumed a log-normal $\lambda$ distribution (for the dark matter)
and we set $\lambda_{\rm crit}$ to be the $f_T(M_\ast)$-th percentile of the distribution, where
$f_T(M_\ast)$ is the \cite{Kelvin+2014} type fraction in that stellar mass bin. The results that
we now discuss are basically identical in these two procedures.

The right-hand panel of Figure \ref{fig:forward_models} shows the comparison of this model with
the observations of \citetalias{RF12}. In this case the model provides a poor representation of
the data in two respects. First, a pure spin-bias scenario, in which ellipticals/spirals form in
low-/high-$\lambda$ haloes at a given mass, can not account for the difference in the average $j_\ast$
at a fixed stellar mass as a function of galaxy type. This is not surprising and can be understood
if one considers that the scatter of the $\lambda$-distribution ($\sim 0.25$ dex) is smaller
than the systematic shift of the Fall relations of discs and ellipticals \citepalias
[$\sim 0.5-0.6$ dex,][]{RF12}.
Second, the shape of the predicted $j_\ast-M_\ast$ relation is non-linear and it steepens at about
the peak mass of $f_\ast$: this results in an inconsistency with the most massive discs and ellipticals
($\log M_\ast/\Msun\gtrsim 11$). Hence we confirm previous claims \citepalias{RF12} that a pure
spin-bias scenario is not able to explain the systematic difference of the specific angular
momentum of spirals and ellipticals of a given stellar mass, even when coupling it with
state-of-the-art SHMR.

Even if we also take into account the fact that elliptical galaxies have typically stopped
forming stars several Gyrs ago, when the specific angular momentum of the halo was smaller
($\jh(z)\propto (1+z)^{-1/2}$, see equation~\ref{eq:j_halo}), we are still not able to explain
all the discrepancy between the relation for spirals and ellipticals in a spin-bias scenario.
This is illustrated by the black arrow in the right-hand panel of Figure~\ref{fig:forward_models},
which shows that even an extreme scenario, in which all ellipticals were already quenched at $z = 3$,
would be at most marginally consistent with the observed shift in normalization with respect to
the spirals\footnote{
Only modest deviations from the SHMR at $z=0$ are expected for that at $z=3$ in the stellar mass
range relevant for ellipticals $M_\ast\gtrsim 10^{10}\Msun$ \citep[e.g.]{Behroozi+2013}.
Hence, for simplicity, in this calculation we have used the SHMR at $z=0$.
}.

Note, also, that our model is extreme, for the transition between spirals and ellipticals is sharp
and unlikely to be realistic. In the case of a smoother transition the two distributions would
move closer and they would be even more inconsistent with the large systematic separation observed
for the Fall relations of spirals and ellipticals.

A pure spin-bias scenario has been recently invoked to explain the formation of the
so-called Ultra Diffuse Galaxies \citep[UDGs, see][]{vanDokkum+2015}, which are very faint
and extended galaxies that are extreme outliers of the mass-size relation. At a given
stellar mass, these systems have on average higher angular momentum than the bulk of the
galaxy distribution \citep[e.g.][]{Leisman+2017} and this has been related to them being
formed in the high-$\lambda$ tail of the halo distribution \citep[][]{AmoriscoLoeb16}.
However, this is not necessarily the case since for a given stellar mass, a galaxy
with an higher-than-average retained fraction $f_j$ would also have an high specific angular
momentum $j_\ast$ for its mass even if the dark matter halo has an average spin
$\lambda$ (see equation~\ref{jstarMstar_theory}). This is also in agreement with results
from recent cosmological simulations, in which the analogous of these galaxies tend to live in
non-peculiar dark matter haloes \citep[see e.g.][]{DiCintio+2017}. In these simulations
stellar feedback prevents the collapse of low-$j$ gas in the central regions and induces
energy and angular momentum exchanges between the dark matter halo and the galaxy, which
ends up having larger angular momentum with respect to other galaxies of similar mass.

\section{Summary and Conclusions}
\label{sec:concl}

In this contribution we have used analytic models in a standard $\Lambda$CDM framework
to study the average retention of angular momentum of galaxies as a function of their mass
and type.
We did this by comparing the stellar specific angular momenta of galaxies on the empirical
Fall relation with that of their dark matter haloes, hence constructing a
\emph{stellar-to-halo angular momentum relation}.
The main improvements with respect to previous works are that we derive the
retained fraction of angular momentum $f_j$ by imposing consistency with both the empirical
Fall relation and the SHMR and that we explore the dependence of the results on the adopted
SHMR. The main results are the following:

\begin{itemize}
\setlength\itemsep{0.5em}
\item we confirm that $f_j$ is smaller than unity for all galaxies and that,
	  on average, late-type galaxies have larger fractions with respect to
      early-types (by about a factor $3-5$) at any given mass;
\item we find a significant dependence of $f_j$ on stellar or halo mass for both morphological
      types and almost all SHMR, with the only exception of the \cite{Dutton+2010} prescription
      for spirals, mostly due to the limited mass range considered there;
\item we find that the $f_j-\Mh$ relation is well represented by a double power-law model
      as in equation \eqref{eq:fits} and we provide the best fit parameters of the double
      power-law function for all the SHMR (Table \ref{tab:fits});
\item we infer a very conservative \emph{upper limit} on the scatter of the $f_j-\Mh$
	  relation of about $0.32-0.36$ dex;
\item we find that the ratio of stellar to halo angular momentum depends on the star formation
      efficiency roughly as $f_j\propto f_\ast^{2/3}$;
\item we confirm that the spin-bias scenario (in which galaxies with different morphologies
      inhabit haloes with different spin) is not sufficient to explain the segregation of
      spirals and ellipticals in the $j_\ast-M_\ast$ plane;
\item our results are consistent with a biased collapse scenario, in which $f_j\propto f_\ast^s$.
      The relatively low value $s\simeq 2/3$ required by the Fall relation is suggestive of
      mixing processes having altered the angular momentum distribution of baryons in the halo
      before or while a fraction $f_\star/f_b$ of them collapsed to form the central galaxy.
\end{itemize}

During the completion of this work, \cite{Shi+2017} published a model of biased collapse to
reproduce the Fall relation and the star formation efficiency. The main differences with our
work are i) their best model predicts a significant flattening of the Fall relation for spirals
at low masses, ii) they do their calculations for only one choice of the SHMR \citep[relatively
similar to the one by][]{Dutton+2010}, iii) they try to disentangle a fraction $f_{\rm inf}$
of accreted material from a fraction $f_j$ of retained angular momentum, based on metallicity
evolution arguments. We refrained from (iii) due to the inherent uncertainties and degeneracy
in the two parameters and, in fact, our $f_j$ includes both processes and should be compared
with their $f_j f_{\rm inf}^s$.

\section*{Acknowledgements}

LP thanks Emmanouil Papastergis and Crescenzo Tortora for many useful discussions.
LP acknowledges financial support from a Vici grant from NWO.
GP acknowledges support from the Swiss National Foundation grant PP00P2\_163824.
EDT acknowledges the support of the Australian Research Council (ARC) through grant DP160100723.

\bsp	
\label{lastpage}
\end{document}